\documentclass[twocolumn,aps,prl,amsmath,amssymb,superscriptaddress,notitlepage]{revtex4-1} 

\usepackage{graphicx}
\usepackage{amssymb}
\usepackage{amsmath}
\usepackage{dsfont}
\usepackage{bm}
\usepackage{mathrsfs}
\usepackage{times}
\usepackage[utf8]{inputenc} 
\usepackage{gensymb} 
\usepackage{xcolor} 

\newcommand{\lp}{\left(}
\newcommand{\rp}{\right)}
\newcommand{\pd}{\partial}
\newcommand{\abs}[1]{\left\lvert #1 \right\rvert}
\newcommand{\fig}[1]{Fig.~\ref{#1}}
\newcommand{\ommin}{\omega_{\rm min}}
\newcommand{\ommax}{\omega_{\rm max}}
\newcommand{\be}{\begin{equation}}
\newcommand{\ee}{\end{equation}}
\newcommand{\eq}[1]{Eq.~(\ref{#1})}
\newcommand{\om}{\omega}

\begin{document}
\title{Observation of noise correlated by the Hawking effect in a water tank}

\author{L.-P.\ Euv\'{e}}
\affiliation{Institut Pprime, UPR 3346, CNRS-Universit\'{e} de Poitiers-ISAE ENSMA 11 Boulevard Marie et Pierre Curie-T\'{e}l\'{e}port 2, BP 30179, 86962 Futuroscope Cedex, France}

\author{F.\ Michel}
\affiliation{Laboratoire de Physique Théorique, CNRS, Univ. Paris-Sud, Université Paris-Saclay, 91405 Orsay, France} 

\author{R.\ Parentani}
\affiliation{Laboratoire de Physique Théorique, CNRS, Univ. Paris-Sud, Université Paris-Saclay, 91405 Orsay, France} 

\author{T.\ G.\ Philbin}
\affiliation{Physics and Astronomy Department, University of Exeter,
Stocker Road, Exeter EX4 4QL, UK}

\author{G.\ Rousseaux}
\affiliation{Institut Pprime, UPR 3346, CNRS-Universit\'{e} de Poitiers-ISAE ENSMA 11 Boulevard Marie et Pierre Curie-T\'{e}l\'{e}port 2, BP 30179, 86962 Futuroscope Cedex, France}

\begin{abstract}
We measured the power spectrum and two-point correlation function for the randomly fluctuating free surface on the downstream side of a stationary flow with a maximum Froude number $F_{\rm max} \approx 0.85$ reached above a localised obstacle. On such a flow the scattering of incident long wavelength modes is analogous to that responsible for black hole radiation (the Hawking effect). Our measurements of the noise show a clear correlation between pairs of modes of opposite energies. We also measure the scattering coefficients by applying the same analysis of correlations to waves produced by a wave maker. 
\end{abstract}

\maketitle

The Hawking effect in laboratory analogues of event horizons~\cite{unr81} has been well studied theoretically~\cite{bar05,cou12} and experiments have been performed in different systems~\cite{rou08,phi08,wei11}. Analogue horizons are created when waves propagate in a stationary counter-flowing medium: at points where the flow speed reaches that of the wave, the latter is blocked and converted to other branches of the dispersion relation. At low frequency, this gives rise to a mode amplification (an over-reflection~\cite{Ache76}) which involves a negative energy wave~\cite{Fabrikant,rou08,PRL09,cou14}, and which is at the root of the Hawking effect~\cite{haw74}. Importantly, the scattered waves of opposite energy are correlated with each other~\cite{Primer}. As a result, when dealing with a noisy system, the two-point correlation function of the fluctuating quantity displays specific patterns both in space-time and in Fourier space~\cite{Carusotto:2008ep,Macher:2009nz,deNova:2015nsa,Busch:2014bza,Boiron:2014npa,ste15}. We here consider surface waves on a stationary counter-current of water in a linear tank. Our work is inspired by the theoretical Refs.~\cite{sch02,PRL09, rou10,mic14} and builds on the experiments~\cite{rou08,wei11,faltot,euv15}. 
As in these experiments, the flow velocity near the blocking point decreases along the direction of the flow. This means that we work with an analog white hole (the time reversed of a black hole). 

\begin{figure}[h]
\includegraphics[width=\linewidth]{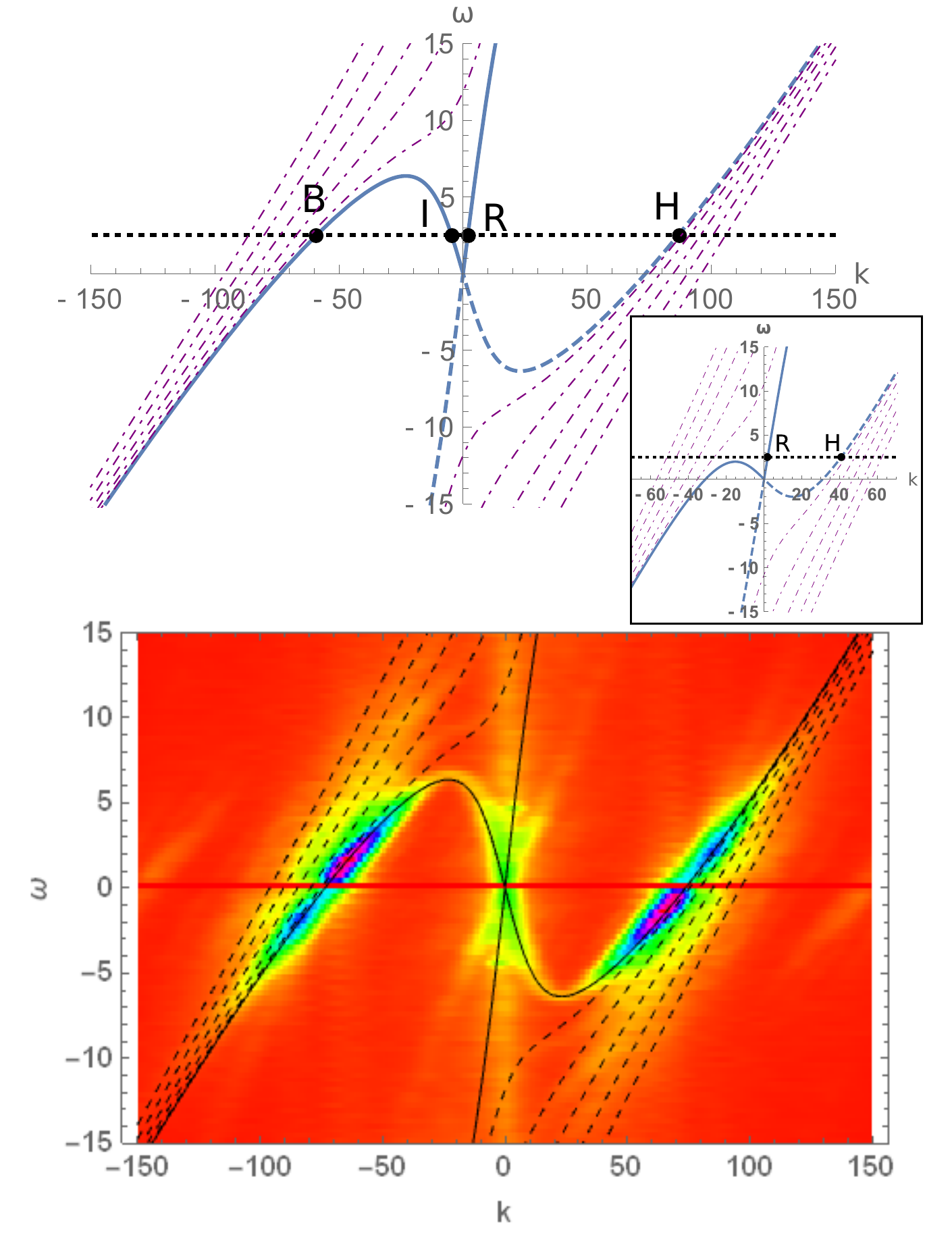} 
\includegraphics[width=0.6 \linewidth]{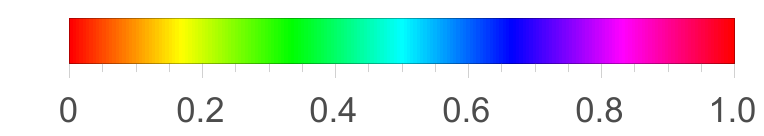}  
\caption{Top: Dispersion relation in the homogeneous flow on the downstream side of the obstacle. $k$ is in ${\rm m^{-1}}$ and the angular frequency $\om$ in Hz. The effective parameters (see text below \eq{PS} for definition) are $U_{\rm eff}=0.37 \, {\rm m.s^{-1}}$ and $h_{\rm eff}=88$~mm. The blue continuous (dashed) lines correspond to \eq{disprel} with positive (negative) $\om - Uk$. 
The four dots labeled by $B, I, R, H$ give the roots $k_a$ for a fixed $\omega > 0$ indicated by a dotted horizontal line. 
Purple, dot-dashed lines describe transverse modes with an even number of nodes in the transverse direction (those with an odd number are not detected by our experimental setup). The inset shows the same dispersion relation for $h = 59 {\rm mm}$, i.e., beyond the turning point for the frequency materialized by the dashed line. Bottom: Square root of the noise power $\mathcal{P}(\omega,k)$ divided by its maximum value and measured on the downstream side of the obstacle, see Fig.~\ref{fig:obstacle} and \eq{PS}.} 
\label{fig:dispersion}
\end{figure}

Ignoring the surface tension, and assuming that the flow is incompressible and irrotational, the dispersion relation which relates the angular frequency $\omega$ and the wave-vector $k$ is
\begin{equation} 
(\omega-Uk)^2=gk\tanh(kh),
\label{disprel}
\end{equation} 
where $U$ is the flow velocity, $h$ the water depth, and $g$ the gravitational acceleration, see the Supplemental material for some explanation about this relation, and its associated wave equation. In a flow to the right, i.e., $U>0$, for a fixed $\omega$, 
see the dotted horizontal line in Fig.~\ref{fig:dispersion}, the three roots $k_I$, $k_B$, and $k_H$ describe counter-propagating waves, i.e., waves with a group velocity oriented to the left in {\it the co-moving frame at rest with the fluid}~\cite{rou08,PRL09,cou12}. Instead $k_R$ describes a co-propagating mode which shall play no role in the sequel. 
There are also transverse modes, which have an effective mass~\cite{jan11} proportional to their transverse wave-vector $k_\perp$. 

\begin{figure}
\includegraphics[width=\linewidth]{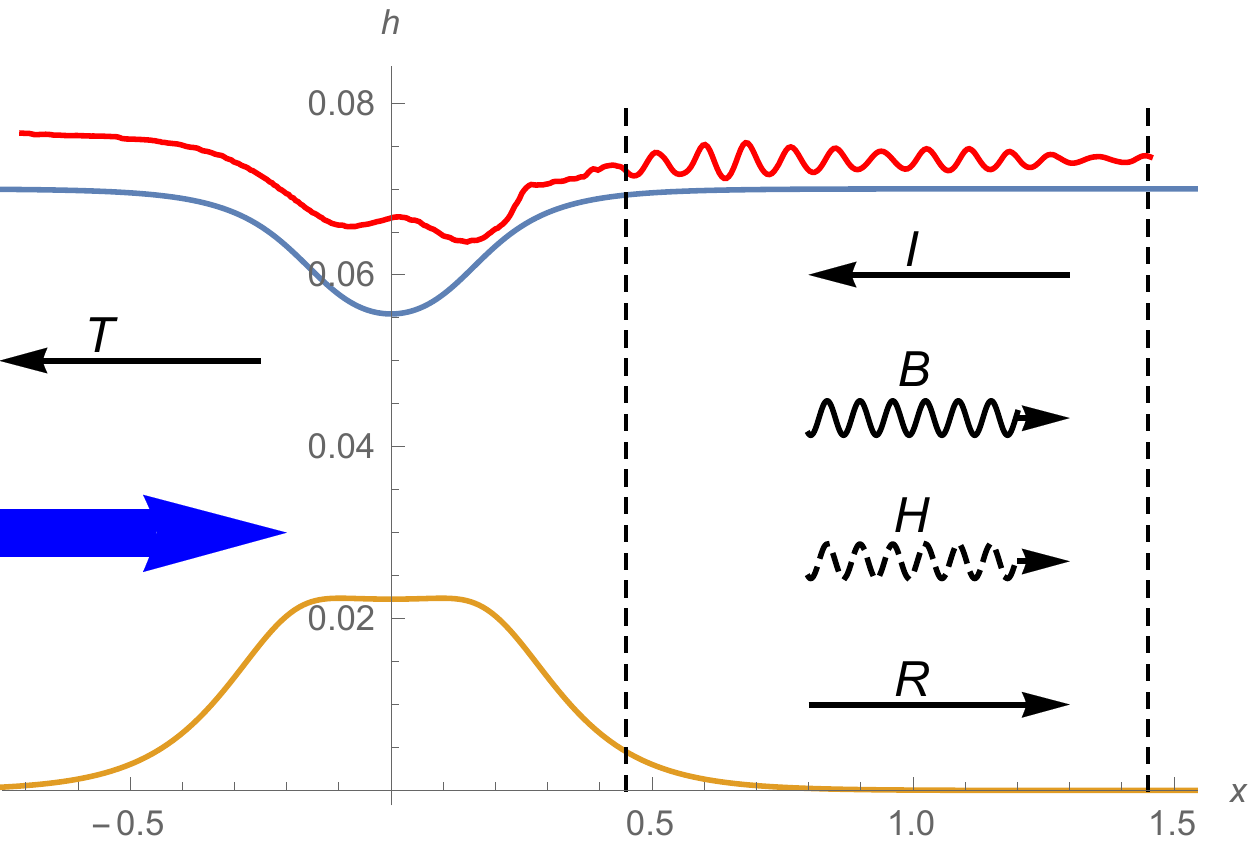}
\caption{Plots of the obstacle (bronze line) and the observed free surface (red line, see also the Figure \ref{fig:zeromode} in the Supplemental Material) in meters. The two dashed vertical lines indicate the region used to study the fluctuations of the free surface $\delta h$. The blue, thick arrow shows the direction of the flow. Thin arrows show the orientation of the group velocity (measured in the laboratory frame) of the various modes produced by the scattering of the incident $I$ mode. The letters $I,B,H,R$ have the same meaning as in Fig.~\ref{fig:dispersion}. The $T$ arrow represents the transmitted wave in the upstream side. The blue curve gives the free surface chosen for determining the obstacle, see text for explanation.} 
\label{fig:obstacle}
\end{figure}

In stationary inhomogeneous flows, such as that of Fig.~\ref{fig:obstacle}, 
$\omega$ is conserved. For fixed $\omega > 0$, the roots $k_I$ and $k_B$ merge at a point in the tank where $U$ becomes sufficiently large~\cite{mic14}. This merging describes an incident long-wavelength mode $I$ coming from the right, that is blue-shifted into a $B$ mode with opposite group velocity $d\omega/dk$ {\it in the laboratory frame} (the slope of the curves in Fig.~\ref{fig:dispersion}): the well-known wave blocking~\cite{Dingemans,PRL09}. We emphasize here that the wave number $k_B$ of the scattered mode is much larger than $k_I$ characterizing the incident wave. This large blue-shifting is the typical signature of analog white hole flows~\cite{Mayo10,cou14}. Importantly, for sufficiently low $\omega$, the wave-blocking is accompanied by a non-adiabatic effect producing an additional mode which also has a large wave vector: $k_H$. This mode has a negative frequency $\omega - U k$ as measured in the fluid frame, see  Fig.~\ref{fig:dispersion}, and thus carries a negative energy~\cite{Fabrikant,rou08,PRL09,cou14}. Because the total wave energy is conserved, this conversion implies an amplification of the $B$ mode. This is in strict analogy with the Hawking effect. However, it is difficult experimentally to have a flow that will block waves at all frequencies and in experiments to date~\cite{rou08,wei11,euv15} only waves above a critical frequency $\omega_\mathrm{min}$ were (essentially) blocked, see the Supplemental material. Above $\omega_\mathrm{min}$, as we shall see, the main effect is the conversion of incident $I$ modes into $B$ and $H$. This effect was reported in~\cite{wei11} both below and above $\omega_\mathrm{min}$. 

\begin{figure}
\includegraphics[width=\linewidth]{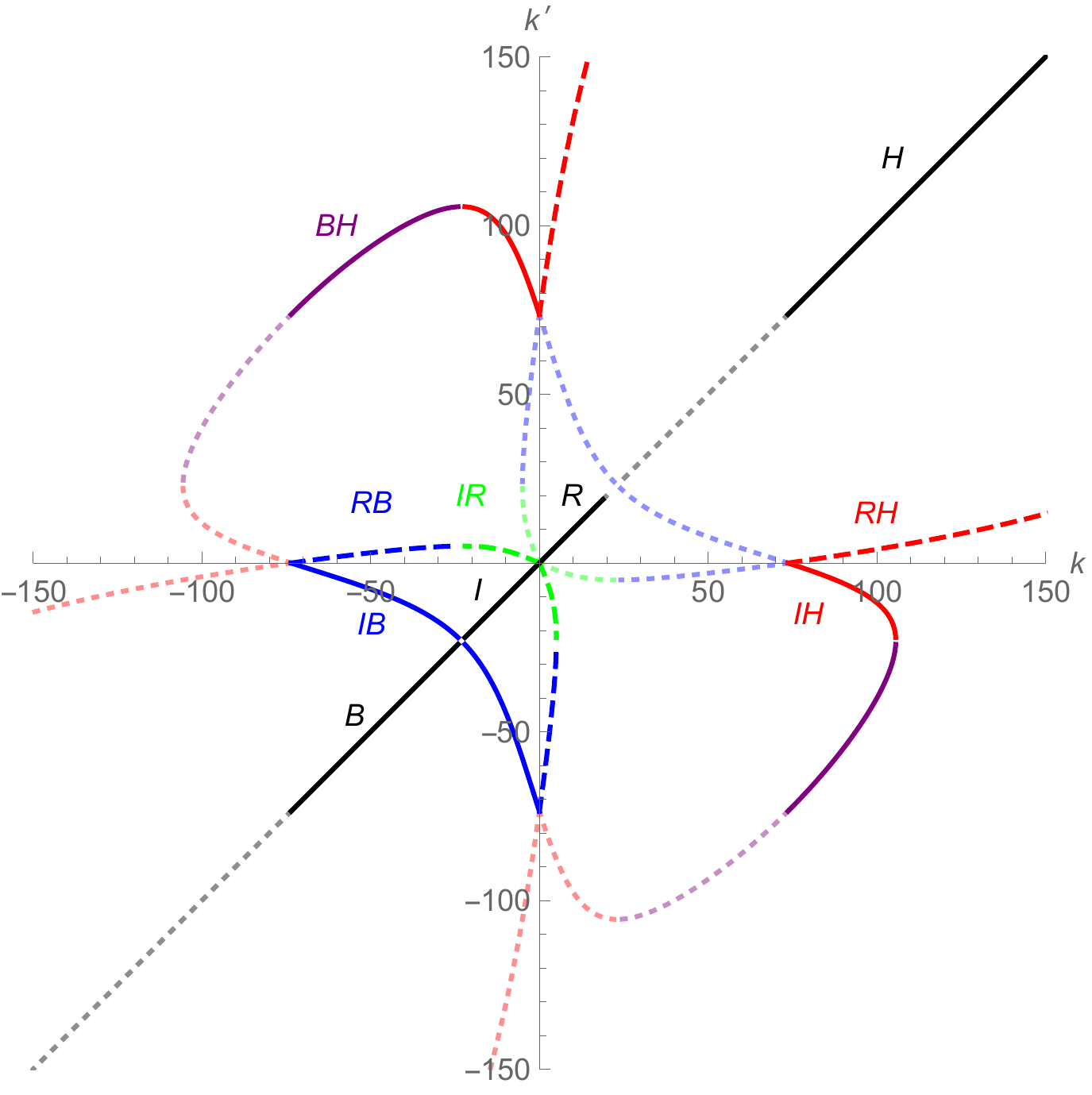} 
\caption{Here we show the loci where $k$, $k'$ are two roots of \eq{disprel} for $\omega \in \mathbb{R}^+$ in a flow with same parameters as in Fig.~\ref{fig:dispersion}. The oblique black segments show $k=k'$ for the four modes $B, I, R, H$. The three continuous curves show $\{k,k'\} = \{k_I,k_B\}$ (blue), $\{k_I,k_H\}$ (red), $\{k_B,k_H\}$ (purple), while the dashed lines $\{k_I,k_R\}$ (green), $\{k_R,k_B\}$ (blue), and $\{k_R,k_H\}$ (red), involve the mode $k_R$. Dotted lighter shaded curves correspond to $\omega < 0$. They are obtained from correlations with positive $\om$ by $(k,k') \to (-k, -k')$.} \label{fig:kpk}
\end{figure}

This conversion can be stimulated by an incident wave $I$ generated by a wave maker, as was done in~\cite{rou08,wei11}. In contrast, the quantum Hawking effect, of fundamental interest for black holes~\cite{haw74}, arises from the amplification of vacuum fluctuations and gives rise to pairs of entangled quanta with opposite energy~\cite{Primer}. Surface waves in the water tank are not suitable to observe the quantum Hawking effect. But just as the quantum vacuum provides the horizon with an irreducible input, there is a stationary background noise of surface waves in the inhomogeneous flow created by both the turbulent flow and the underwater obstacle, see Fig.~\ref{fig:dispersion} lower panel. Because of the mode conversion near the blocking point, this noise should be {\it correlated}. When measured in the downstream homogeneous region, see Fig.~\ref{fig:obstacle}, these correlations are non-vanishing when the $k_a$'s are evaluated at the same value of $\omega$~\cite{Macher:2009nz,deNova:2015nsa,Boiron:2014npa}, see Fig.~\ref{fig:kpk} and the Supplement material. 

Our experiments were performed in the water channel of the Pprime Institute (for more details, see Supplemental Material). The obstacle used to obtain an inhomogeneous flow was designed following the procedure outlined in appendix A of~\cite{mic14}. It relies on the hodograph transformation for a 2D inviscid, irrotational, incompressible flow~\cite{unr13}. The shape of the obstacle is determined by the profile of the free surface, the asymptotic water depth, and flow velocity, see Fig.~\ref{fig:obstacle}. The main advantage of this obstacle over the one used in~\cite{wei11,euv15} is that it supports a flow with a relatively large Froude number: $0.86 \pm 0.03$ in the present experiment instead of $0.67\pm 0.02$ in~\cite{euv15}. In addition, it produces a smaller static surface deformation, or undulation~\cite{cou14,unr08}, with a peak-to-peak amplitude of a few millimeters (see Supplemental Material). The descending slope of the obstacle also has a larger maximum gradient: the slope of $c - U$, giving the analogue surface gravity in transcritical flows~\cite{mic14},  has a maximum of $2$~Hz instead of $1.2$~Hz as used in~\cite{wei11} (here $c = \sqrt{g h}$ is the velocity of long-wavelength waves in the fluid frame).

We measured the fluctuations of the water height $\delta h(x,t)$, defined as the deviation from the time-averaged value of $h(x,t)$, in the downstream constant-flow region shown in Fig.~\ref{fig:obstacle}. We first studied the noise power (which is proportional to the wave action \cite{bg68}) defined by 
\begin{equation} 
\label{PS}
\mathcal{P}(\omega,k) \equiv \left\langle \left\lvert \delta \tilde{h}(\omega,k) \right\rvert^2 \right\rangle \times S_k^{-2}. 
\end{equation}
Here, $\delta \tilde{h}(\omega,k)$ is the Fourier transform of $\delta h(t,x)$ and $S_k = |gk\tanh(kh)|^{1/4}$ is the structure factor relating plane waves to unit norm modes when working at fixed $k$~\cite{cou14,mic14}. The Fourier transform in time is computed using a rectangular window, while we used a Hamming window function~\cite{Ham} with support $x \in [0.45 {\rm m}, 1.45 {\rm m}]$ to compute the spatial transform (see Supplemental Material). The mean value is computed by dividing the data into 80 pieces of equal duration ($12.5$~s) and averaging over them. In former studies of the noise~\cite{wei13,faltot}, this averaging was not performed. As a result the plots showed random values of $| \delta \tilde{h}(\omega,k) |^2$ as opposed to its mean. The square root of $\mathcal{P}(\omega,k)$ is shown in the lower panel of Fig.~\ref{fig:dispersion}. 

Although the upstream water height and flow velocity were $h_{\rm up} =74\ {\rm mm}$ and $U_{\rm up} = 0.31 \, {\rm m s^{-1}}$, the dispersion relation of Fig.~\ref{fig:dispersion} has been drawn with the effective values $h_{\rm eff} = 88$~mm and $U_{\rm eff}=0.37 {\rm m s^{-1}}$, chosen to match the observed wave numbers. The agreement of the dispersion relation with the three counter-propagating modes $I,\, B$, and $H$ is clear for all values of $\omega$. We expect that the differences with $h_{\rm up}$ and $U_{\rm up}$ are due to boundary-layer, vorticity, and turbulent effects. When using $h_{\rm eff}$ and $U_{\rm eff}$, we find that the value of $\omega$ for which the two roots $k_I$ and $k_B$ merge on top of the obstacle (respectively in the downstream asymptotic region) is $\omega_{\rm min} \approx 0.8$~Hz (respectively $\omega_{\rm max} \approx 5$~Hz). 

\begin{figure} 
\includegraphics[width=0.49 \linewidth]{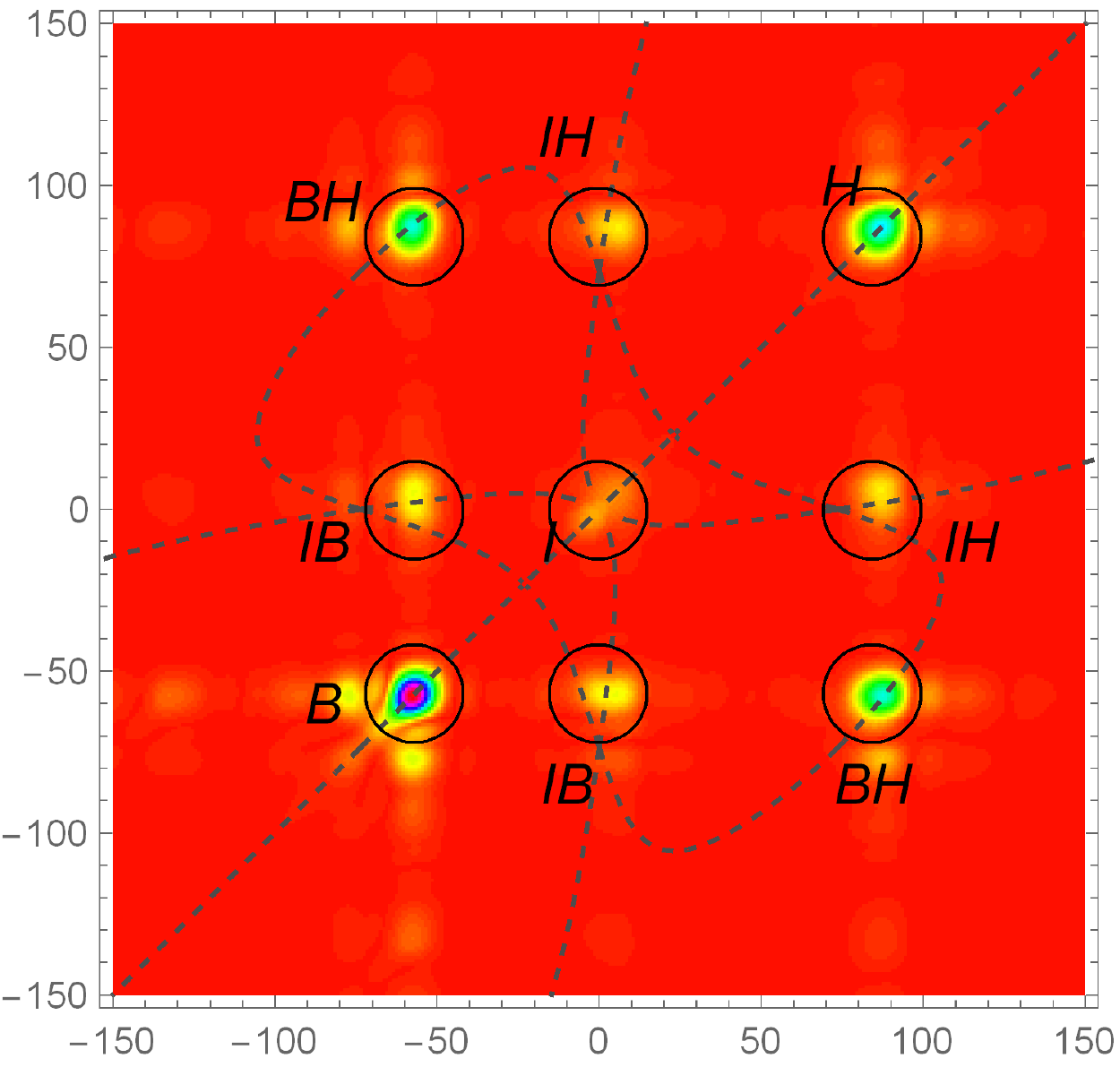}
\includegraphics[width=0.49 \linewidth]{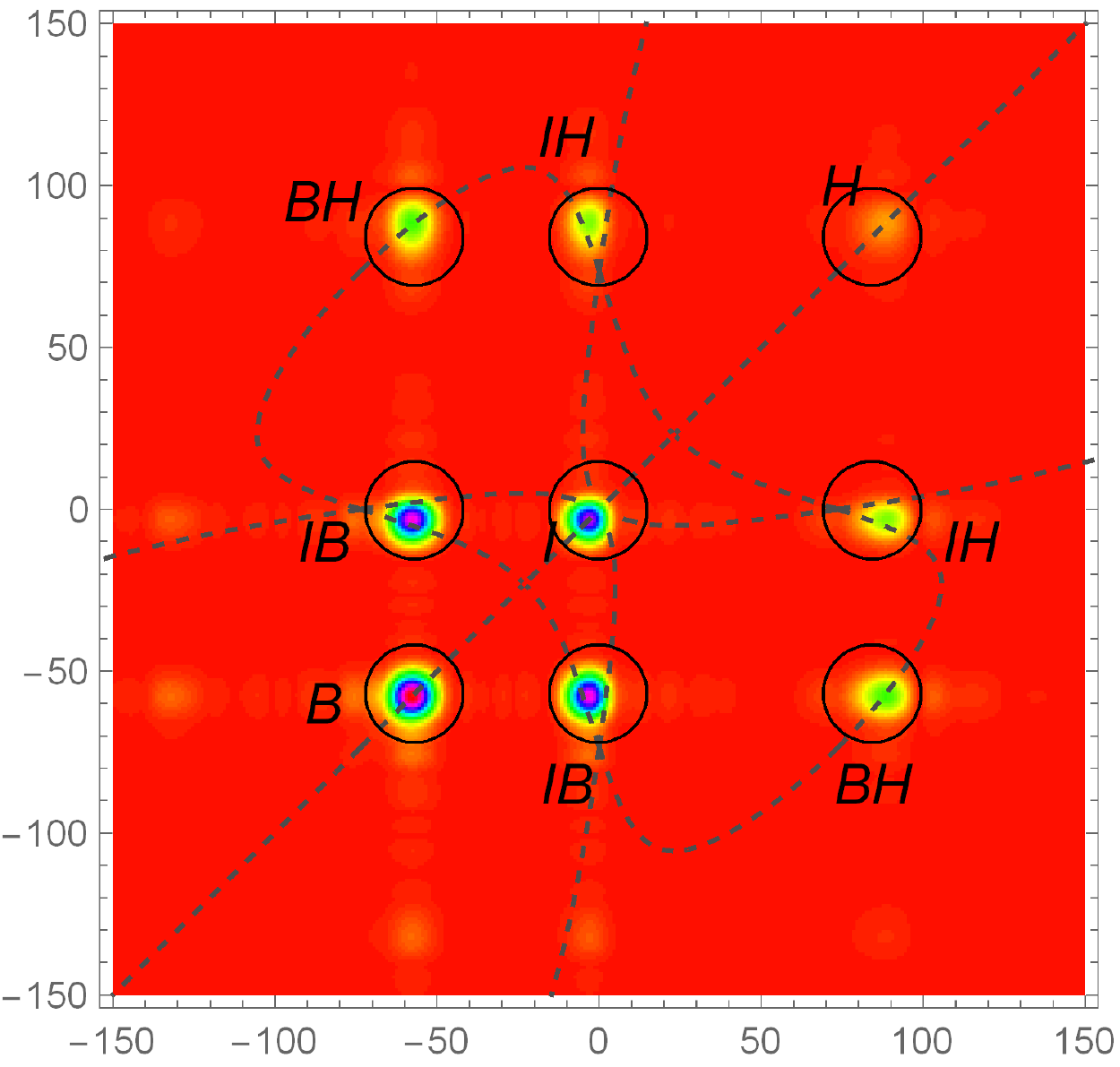}
\caption{In the left panel, we show the noise correlation function of Eq.~(\ref{eq:G2}) for $\omega = 2.5$~Hz. 
The color scale is the same as in Fig.~\ref{fig:dispersion}. Dashed lines show the dispersion relation in the $(k,k')$ plane, see Fig.~\ref{fig:kpk}. The circles are centred on $(k_\omega,k'_\omega)$ where $k_\omega$ and $k'_\omega$ are two roots of the dispersion relation for the considered frequency $\omega$. The letters $B$ and $H$ designate the power of the short wave length modes with opposite energies, and $BH$ their correlations. 
In the right panel, we show again Eq.~(\ref{eq:G2}) when the wave-maker is sending the incident wave $I$ for the corresponding frequencies. The $IB$ and $IH$ correlations are clearly visible.}\label{fig:G2nwm} 
\end{figure}
We then measured the two-point correlation function evaluated at the same frequency and two different wave-vectors:
\begin{equation}\label{eq:G2}
G_2(\omega ; k,k') \equiv \left\lvert \left\langle \delta \tilde{h}(\omega,k) \delta \tilde{h}(\omega,k')^* \right\rangle \right\rvert \times (S_k S_{k'})^{-1}.
\end{equation}
In the left plot of Fig.~\ref{fig:G2nwm}, we show $G_2(\omega ; k,k')$ in the $(k,k')$ plane for $\omega = 2.5$~Hz. 
Using the effective values $h_{\rm eff}$ and $U_{\rm eff}$, the non-vanishing correlations in the $k,k'$ plane are found along the lines drawn in Fig.~\ref{fig:kpk}, as expected. 
We note that the $B$ and $H$ modes of opposite energy are well correlated. 
We also note that the long-wavelength modes $I$ and $R$ are correlated with both $B$ and $H$ modes. However, we cannot clearly separate the contributions of $I$ and $R$ modes. 
Since numerical simulations (see Supplemental material) indicate that the co-propagating ($R$) mode is only weakly coupled to $I, B$ and $H$, in what follows, we only study the power and the strength of correlations of these three modes.

\begin{figure} 
\includegraphics[width=\linewidth]{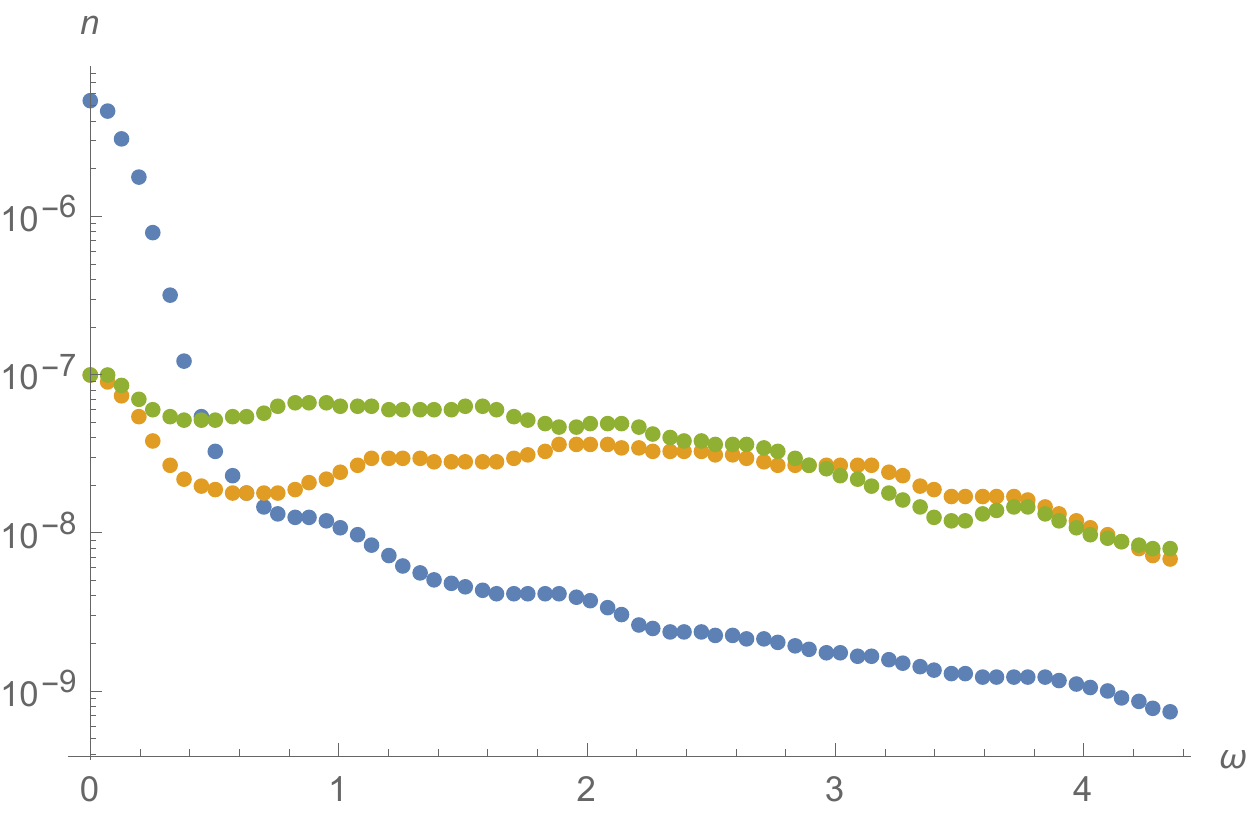}
\includegraphics[width=\linewidth]{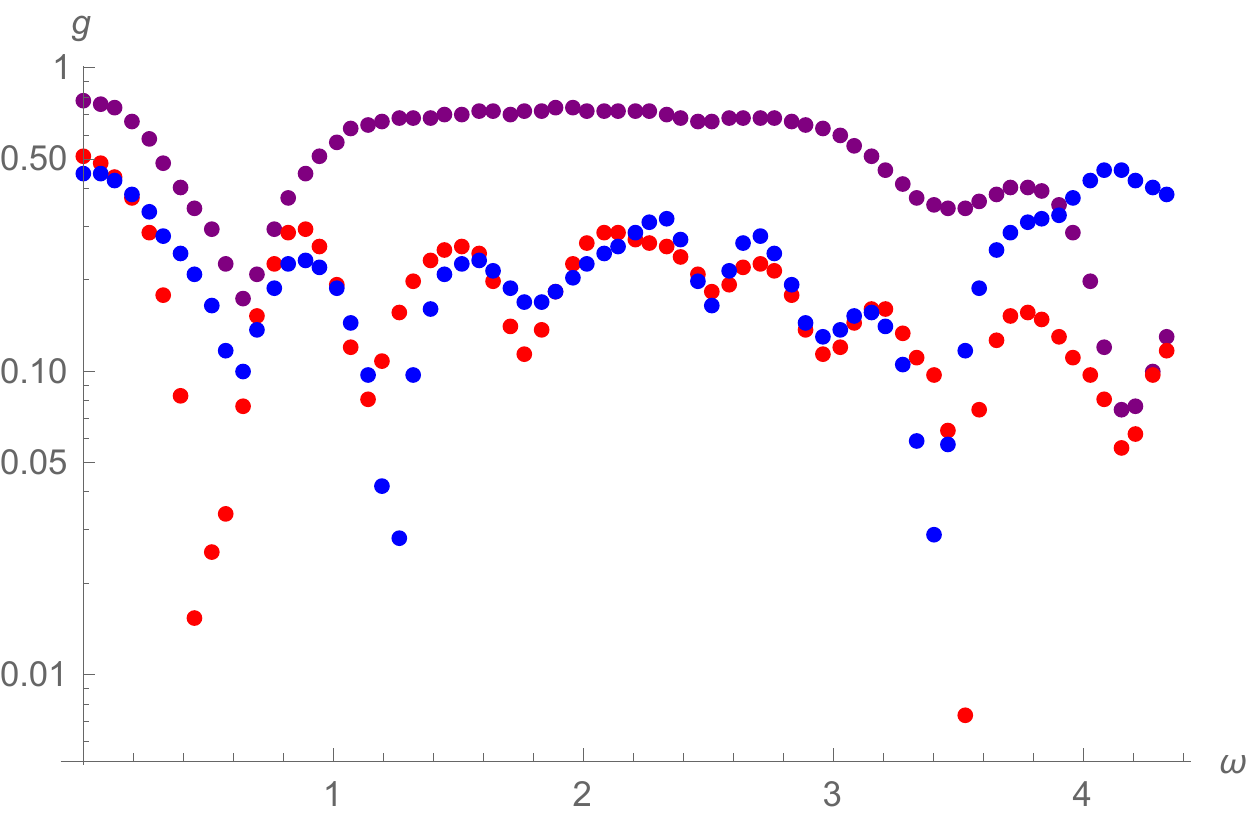}
\caption{The noise power $n_a(\omega)$ (upper plot, in ${\rm m^3.s^{-2}}$) and the relative strength of the correlations of Eq.~(\ref{eq:g2om}) (lower plot), in logarithmic scale, for the counter-propagating modes $I,B$ and $H$. In the upper plot, the power of $I$ is in blue, that of $B$ in green, and in bronze that of $H$. In the lower plot, the correlations $BH$ are in purple, $IB$ in blue, and $IH$ in red. We estimate that the uncertainties are of order $0.1$.
}\label{fig:cij} 
\end{figure}

In the upper plot of Figure~\ref{fig:cij} we show $n_a(\omega) \equiv \mathcal{P}(\omega,k_a)/|dk_a/d\om|$ as a function of $\omega$,
where $k_a(\omega)$ are the three counter-propagating roots. This quantity gives, up to an overall factor, the {mean} number of quasi-particles per unit angular frequency interval~\cite{cou14,mic14}. The power spectra of the two dispersive modes $B$ and $H$ are comparable except in a domain near $\omega_{\rm min} \approx 0.8$~Hz where there are more $B$-modes. The hydrodynamical $I$-modes have less power by a factor $\sim 10$, except below $\omega_{\rm min}$ where their power is much larger. (In the absence of obstacle, the observed noise power is completely dominated by the hydrodynamical modes $I$ and $R$, and there are no significant correlations between $I$ and $B, H$ modes, see Supplemental material.) 

To quantify the strength of the correlations, we study the ratio of the cross-correlations over the square root of the product of the auto-correlations 
\begin{equation} \label{eq:g2om}
g_2(\omega ; a,b) \equiv \frac{G_2(\omega ; k_a,k_b)}{\sqrt{G_2(\omega ; k_a,k_a) G_2(\omega ; k_b,k_b)}}, 
\end{equation} 
where $k_a(\omega), k_b(\omega)$ are two roots for a given $\om$. 
For any statistical ensemble, $g_2$ is necessarily smaller than 1. As explained in the Supplemental material, $g_2$ involves the classical counterparts of the observables which are currently used in quantum settings to assert that the state of the scattered waves $k_a(\omega), k_b(\omega)$ is entangled~\cite{deNova:2015nsa,Busch:2014bza,Boiron:2014npa,ste15}.  
The lower plot of Figure~\ref{fig:cij} shows three types of correlations: the $BH$ correlations are stronger than the two other ones, since $g_2(\omega; B, H)$ is close to $0.7$ (except near $0.6$~Hz). This indicates that $70\%$ of $B$ and $H$ modes are in correlated $BH$ pairs. The $IH$ and $IB$ correlations are below $0.3$ over most of the frequency domain. This implies that more than $50 \%$ of $BH$ pairs do not come from observed $I$ modes with $k_\perp = 0$.
It probably means that a significant fraction of $BH$ pairs have a non-vanishing $k_\perp$. 
(At present we are not able to separate the contributions of $B$ and $H$ modes with and without transverse wave number, as the corresponding curves on the dispersion relation are very close to each other, see Fig.~\ref{fig:dispersion}.) Some $BH$ pairs should also be produced by {\it incident} waves $H$ and $R$ from the left. 
In addition, not all the incident $I$-mode noise is taken into account if some of it is generated by fluctuations in the region $x < 0.45$m.  
An effective description of this generation could be obtained from adapting to the present case the driven-damped wave equation of Ref.~\cite{scot15}. 

The properties of the scattering can be more clearly studied when sending a $I$ wave towards the obstacle, as was done in~\cite{rou08,wei11}. The corresponding values of $G_2$ are shown in the right panel of Fig.~\ref{fig:G2nwm} 
for $\omega \approx 2.5 {\rm Hz} > \omega_{\rm min}$. The power of the reflected $B$ wave is close to that of the incident one, as expected from the validity of the WKB approximation in this regime~\cite{cou12}. The negative-energy $H$ wave remains relatively small in amplitude. However, $HI$ and $HB$ correlations are clearly visible, showing that $H$ and $B$ waves are produced by the analog Hawking effect. This is further clarified by the analysis of the scattering coefficients presented in the Supplemental Material. 

To summarize, we observed the statistical properties of the water depth fluctuations downstream from an obstacle in a flow with a large maximum Froude number.  The negative energy modes $H$ are highly populated, strongly correlated with the positive energy modes $B$, but more weakly correlated with the $I$ modes. The noise correlations have the main features expected from the Hawking effect, whose correlations we observed also in the stimulated case with a wave maker. Further experiments and theoretical work are required to clarify all of the processes behind these observations. 

{\it Acknowledgements} : We thank Y. Stepanyants for comments. We acknowledge support from the University of Poitiers (ACI UP on Wave-Current Interactions 2013-2014), the Interdisciplinary Mission of CNRS (PEPS PTI 2014 DEMRATNOS), the University of Tours (ARC Poitiers-Tours 2014-2015), the French national research agency (ANR) HARALAB (N\degree ANR-15-CE30-0017-04), the FEDER 35790-2012, and a FQXi grant of the Silicon Valley Community Foundation.

\vspace*{1 cm}

\newpage

\section{Supplemental Material}
\setcounter{equation}{0} 
\renewcommand{\theequation}{S\arabic{equation}} 
\setcounter{figure}{0} 
\renewcommand{\thefigure}{S\arabic{figure}} 

\subsection{Dispersion relation and critical frequencies in inhomogeneous flows} 

To describe theoretically the system used in the experiment, we assume water in the flume can be approximated by an ideal, incompressible fluid in laminar, irrotational motion. We also temporarily assume the velocity component along the transverse direction of the flume is negligible, so as to work with an effectively two-dimensional flow. Under these approximations, the linear fluid equations describing the evolution of small perturbations can be integrated from the bottom up to obtain an effective one-dimensional equation involving only quantities defined at the free surface, see~\cite{Unruh:2013gga, Coutant:2012mf} for the detailed derivation. Assuming further that the vertical component of the velocity is negligible before the horizontal one (which is satisfied far from  the obstacle), this equation takes the form
\begin{align}\label{eq:UCP1}
 & \left[\left( \partial_t + \partial_x U(x) \right) \left( \partial_t + U(x) \partial_x \right) 
 \right. \nonumber \\ & \left.
 \hspace*{1 cm} + i g \partial_x  \tanh \left( i h(x) \partial_x \right) \right] \delta \phi(t,x) = 0,
\end{align}
where $\delta \phi$ denotes the perturbation of the velocity potential (i.e., $\partial_x \delta \phi$ gives the perturbation of the velocity), $h$ is the local water depth of the background flow, and $U$ its velocity evaluated at the free surface. 
We used a quartic expansion of this equation in the numerical simulations reported in the following. 
The total water height $h_T$ is given by the sum the background height and the time-dependent linear fluctuation 
\be \label{eq:dh}
h_T(t,x) = h(x) -\frac{1}{g}\left(\partial_t+ U \partial_x\right)\delta \phi. 
\ee

In a region where $U$ and $h$ are homogeneous, \eq{eq:UCP1} becomes
\begin{align}
\left[ D_t^2 + i g \partial_x \tanh \left( i h \partial_x \right) \right] \delta \phi(t,x) = 0,
\end{align}
where $D_t \equiv \partial_t + U \partial_x$ is the time derivative in the rest frame of the fluid. 
Since the $x$ direction plays no particular role in this frame (except for the boundary conditions), one can relax the assumption that $\delta \phi$ be independent on $y$. 
Doing so, we obtain
\begin{align}\label{eq:UCP2}
\left[ D_t^2 + i g \vec{\nabla} \tanh \left( i h \vec{\nabla} \right) \right] \delta \phi(t,x,y) = 0,
\end{align} 
where $\vec{\nabla} = \binom{\partial_x}{\partial_y}$. One can look for plane wave solutions in 
\begin{align}
\delta \phi(t,x) \propto \exp \left( i \left( - \omega t + k x \right) \right) \cos \left( n \pi y / l \right),
\end{align}
where $y$ is a Cartesian coordinate in the transverse direction, $l$ is the width of the flume, $n \in \mathbb{N}$, and $(\omega, k) \in \mathbb{R}^2$. 
The last factor comes from the boundary conditions at the walls of the flume, i.e., that the $y$-component of the velocity, given by $\partial_y \delta \phi$, must vanish at $y = 0$ and $y = l$.  
(For definiteness, we set the origin of $y$ on one wall of the flume.) 
Plugging this into \eq{eq:UCP2} and noticing that, since $\tanh$ is an odd function, the second term involves only even-order derivatives in each direction, give the dispersion relation
\begin{align}
\left( \omega - U \cdot k 
\right)^2 = g p(k,n) \tanh \left( h p(k,n) \right),
\end{align}
where $p(k,n) \equiv \sqrt{k^2 + \frac{n^2 \pi^2}{l^2}}$ is the modulus of the projection of the wave vector on the free surface. 
Taking the small-$k$ limit, one can see that the velocity of long-wavelength waves in the fluid frame is $c = \sqrt{g h}$. The Froude number $F$, defined as the local fluid velocity divided by $c$, is thus equal to $U / \sqrt{g h}$. 
Modes with $n = 0$ (continuous and dashed lines in the top panel of \fig{fig:dispersion}) are homogeneous along the transverse direction, while those with $n > 0$ are the transverse modes (dot-dashed lines in the Figure). 
For the former, we distinguished modes with positive (continuous line) and negative (dashed line) frequency $\omega - U k$ in the rest frame of the fluid. 

At fixed $\om$, the dispersion relation of longitudinal modes, see \eq{disprel}, has an infinite number of roots in $k$, but only a few of them are real. More precisely, if the flow is subcritical, i.e., if $\left\lvert F \right\rvert < 1$, there are 4 real roots for $\abs{\omega}$ smaller than a critical value $\om_c(v,h)$, and two real roots otherwise: two roots, corresponding to the $I$ and $B$ modes, merge and become complex when $\om$ crosses $\om_c$ by increasing values. The value of $\om_c$ depends 
on $U$ and $h$, which leads us to define two important frequencies, see \fig{fig:maxmin}:
\begin{figure} 
\centering
\includegraphics[width =  0.23 \textwidth]{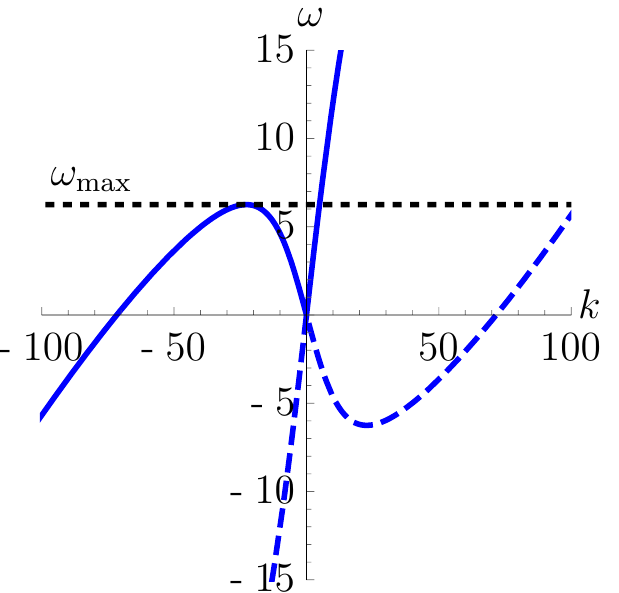}
\includegraphics[width =  0.23 \textwidth]{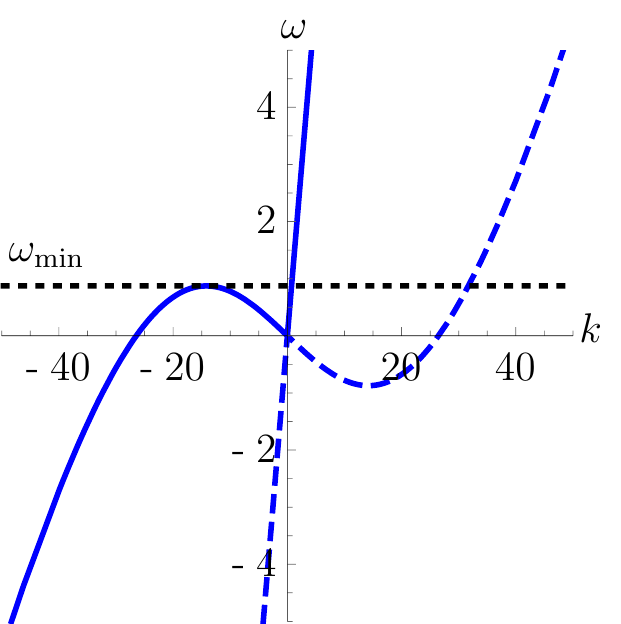}
\caption{Dispersion relation for the longitudinal modes ($n = 0$), evaluated in the downstream asymptotic region (left panel) and at the point where the water depth reaches its minimum (right panel). 
The horizontal, dotted lines show the corresponding values of $\om_c$ where the $I$ and $B$ roots merge, i.e., $\om_{\rm max}$ on the left panel and $\om_{\rm min}$ on the right one. 
For the flow we used, see \fig{fig:obstacle}, one has $\om_{\rm max}\approx 5$Hz and $\om_{\rm min} \approx 0.8$Hz. 
}
\label{fig:maxmin}
\end{figure} 
\begin{itemize}
\item $\omega_{\rm max}$ is the $x \to \infty$ limit of $\om_c(h(x), v(x))$. It is thus the maximum value at which the wave-maker can send a counter-propagating wave. 
\item $\omega_{\rm min}$ is the minimum value of $\om_c$, reached close to the top of the obstacle. In the WKB approximation, an incoming counter-propagating wave with $\om < \om_{\rm min}$ has no turning point and is thus essentially transmitted across the obstacle. On the contrary, a wave with $\om > \om_{\rm min}$ has a turning point and will be essentially reflected.
\end{itemize} 
If the flow is (locally) supercritical, i.e., $\abs{F} > 1$, then there are only two real roots for any real value of $\omega$. 
So, in a transcritical flow (where, by definition, there is a supercritical region), $\om_{\rm min} = 0$. 
In our setup instead, the Froude number is everywhere smaller than $1$, so $\om_{\rm max} > \om_{\rm min} > 0$. Their respective values are $\om_{\rm max}\approx 5$Hz and $\om_{\rm min} \approx 0.8$Hz. 

In a stationary background flow, the frequency of linear perturbations is conserved. The outgoing waves obtained by the scattering of an incoming monochromatic wave of frequency $\om$ thus have the same frequency as the incoming one. Hence their wave vectors $k^a$ evaluated far away from the obstacle will be given by the real roots of \eq{disprel} at that frequency $\om$. 
One consequence of this is that 2-point correlation function of the noise in the $(k,k')$ plane is localized where $k$ and $k'$ correspond to the same frequency. This is shown in \fig{fig:kpk}. 
To obtain the curves, we considered all the possible doublets $(k_a(\om), k_b(\om))$ where $(a,b) \in \left\lbrace B, I, R, H \right\rbrace^2$, and drew their locus when $\omega$ describes $\left[- \omega_{\rm max}, \omega_{\rm max} \right]$. These curves are also shown in \fig{fig:G2nwm} to verify that the location of the peaks is consistent with our analysis, and to identify the waves they involve. 

\subsection{Experimental Setup} 

Our experiments were performed in the water channel of the Pprime Institute, which is 6.8~m long and 0.39~m wide. A PCM Moineau pump creates the water current and its flow rate is regulated by a variator. The water is first injected into a convergent chamber whose geometry and honeycomb structure produce an outgoing flow devoid of practically all boundary-layer effects and macro-vortices. At the downstream end of the channel, there is a guillotine which can be set in vertical motion around a mean height to generate a wave that propagates upstream, see Fig.~1 in \cite{euv15}. To measure the free surface, as in~\cite{wei11,faltot}, a laser sheet was projected from above along the center line of the channel.  
The laser beam is produced by an Argon LASER (Spectra Physics 2W), guided by an optical fiber to a cylindrical lens (placed at $1.44$ m above the free surface) which formed a LASER sheet. With a LASER power of $0.14$ W (after the lens), the power density at the free surface is around $0.05\ {\rm Wm^{-1}}$. A fluorescent dye was added to the water to delineate a laser line on the surface, specifically $50$ g of fluorescein to reach a high concentration ($\approx 12\ {\rm gm^{-3}}$) in order to obtain a minimum penetration of the LASER sheet into the fluid. Indeed, the LASER light intensity decreases by $90\%$ at $5$ mm below the free surface. Three cameras (Jai CVM2 1600x1200) captured this LASER line on a $2.16\ {\rm m}$ wide visualization window. Their resolution is $0.45\ {\rm mm/px}$. We used a sub-pixel detection method similar to that of~\cite{wei11} with an accuracy of $0.1\ {\rm mm}$, see also~\cite{faltot}. The Figure \ref{fig:zeromode} shows the stationary undulation (represented after free surface detection with sub-pixel accuracy in Figure~\ref{fig:obstacle} by the red curve). 

\begin{figure*}
\includegraphics[width=0.9 \textwidth]{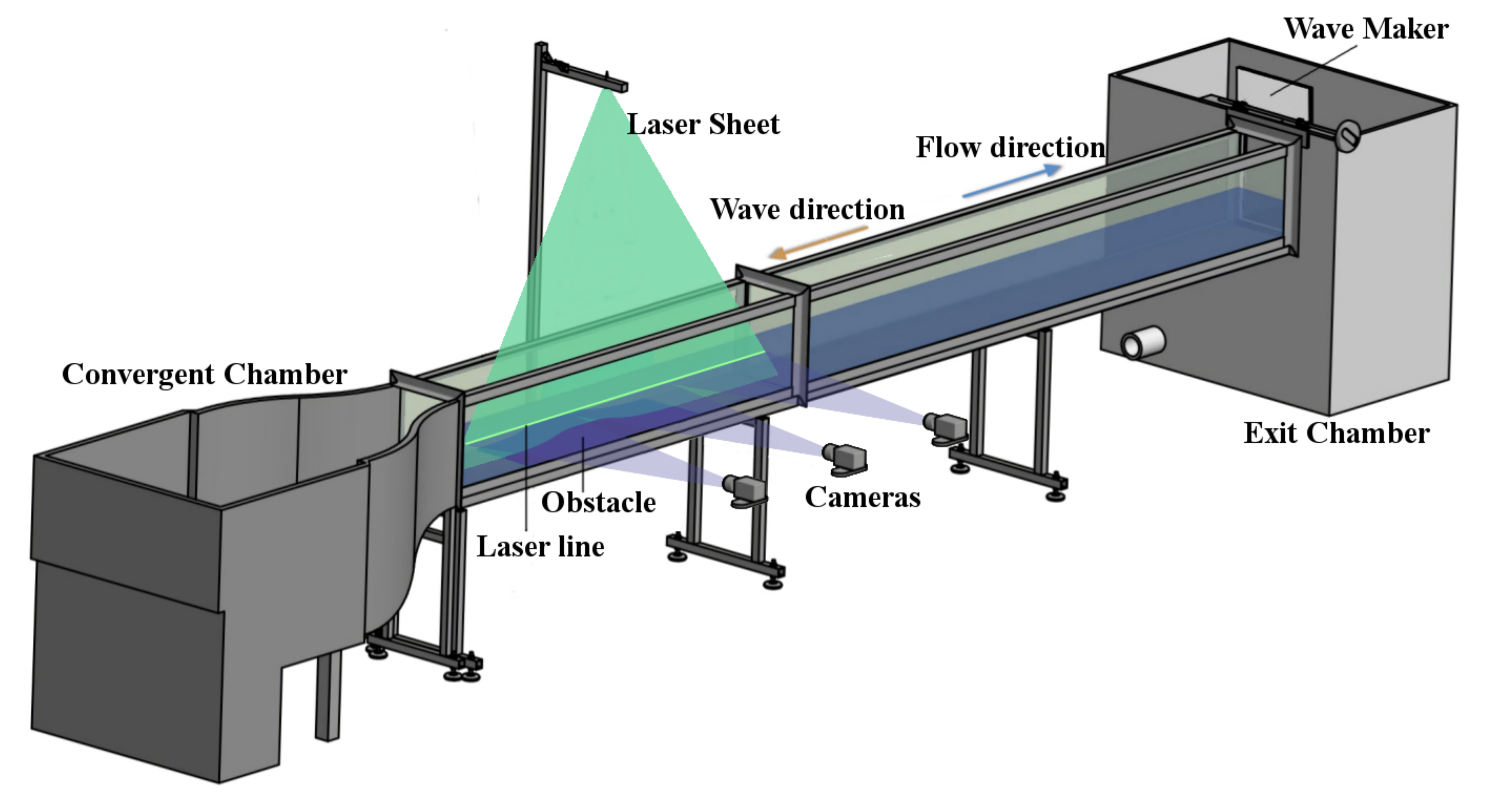}
\caption{A sketch of the experimental setup for measuring free surface deformation. The flow goes from left to right, as indicated by the blue arrow, while the incident mode $I$ sent by the wave maker goes in the opposite direction, see brown arrow. 
}\label{fig:canal}
\end{figure*}

\begin{figure*}
\includegraphics[width=0.9 \textwidth]{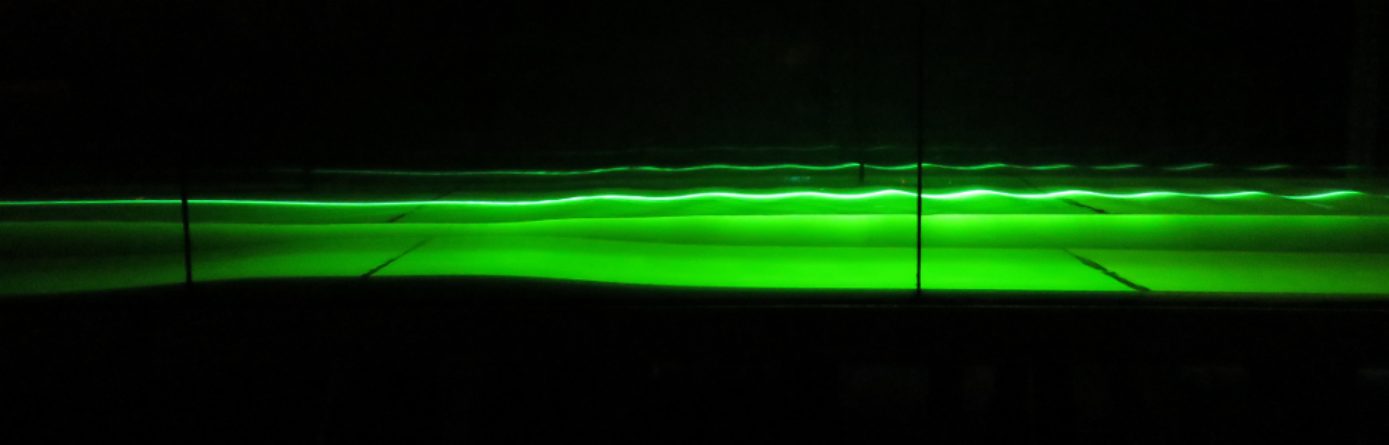}
\caption{Picture of the LASER line showing the stationary undulation over the obstacle. The distance between both vertical black lines is $1\ {\rm m}$.
}\label{fig:zeromode}
\end{figure*}

\subsection{Windowing} 

\begin{figure*}
\includegraphics[width= 0.49 \linewidth]{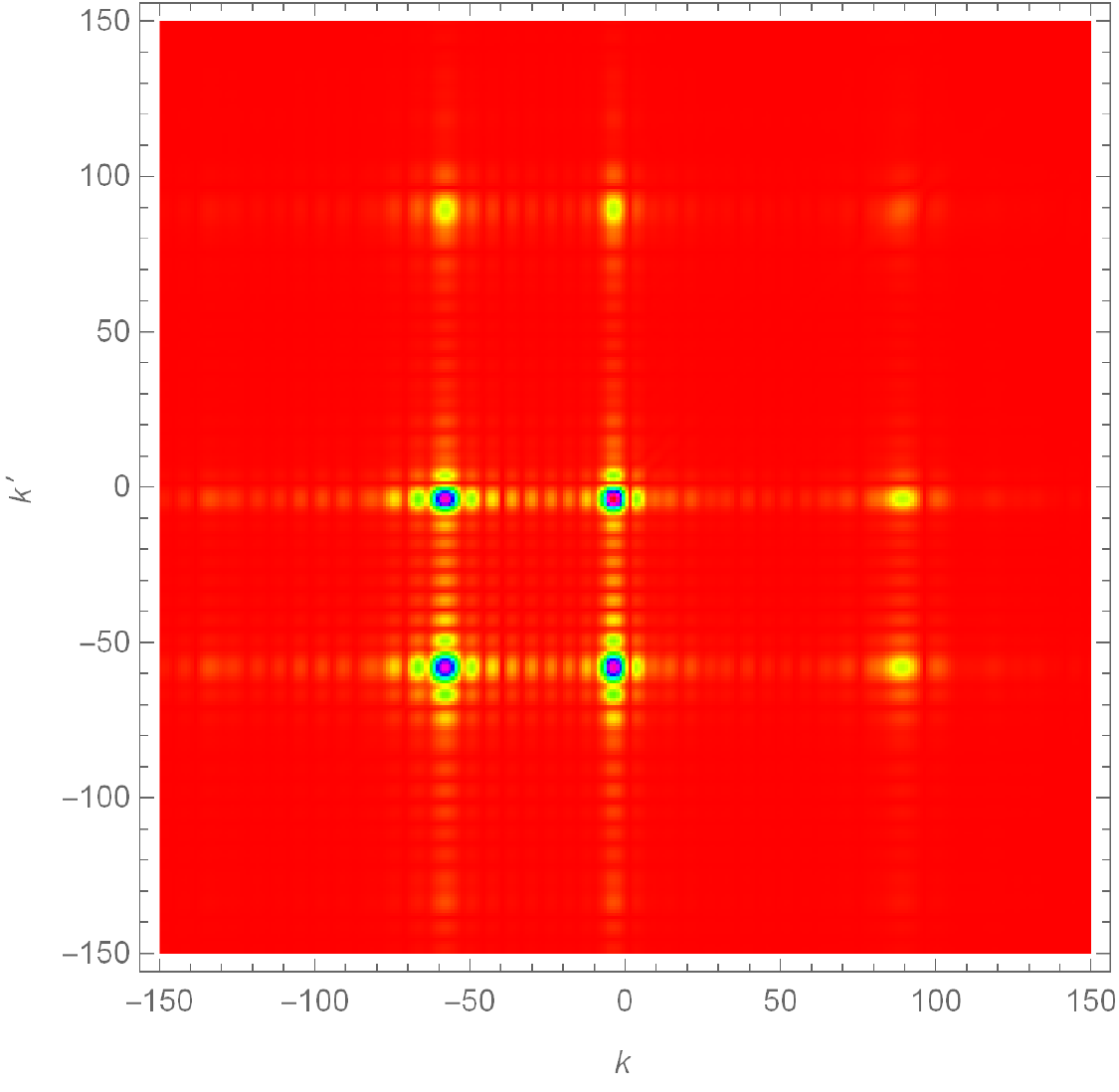}
\includegraphics[width= 0.49 \linewidth]{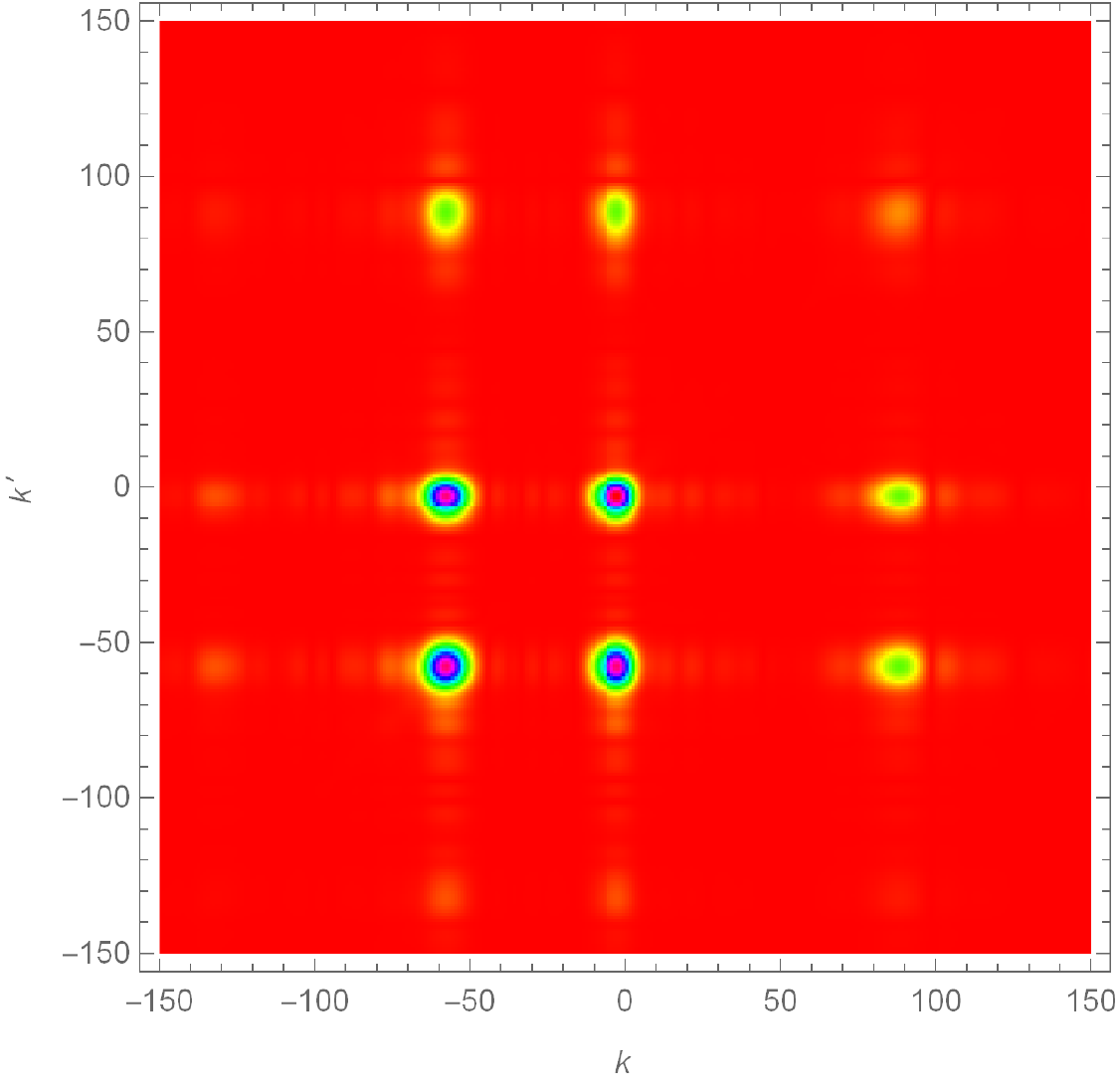}
\caption{Comparison of the two-point correlation function $G_2(\omega;k,k')$ of \eq{eq:G2} 
evaluated using a rectangular window (left) and a Hamming window (right) over $x \in [0.45 {\rm m}, 1.45 {\rm m}]$, for the case when an incident wave of angular frequency $\om = 2.5$ Hz. $k$ and $k'$ are expressed in ${\rm m^{-1}}$.}\label{fig:window}
\end{figure*}

The finite length $L_I$ of the spatial integration window has two undesirable effects. First, the Fourier transform of a plane wave has a main lobe with a non-vanishing width, which limits the accuracy in determining the wave-vector $k$ and amplitude $A$. Second, it also shows side-lobes which produce artefacts in the correlation functions. At fixed $L_I$, the magnitude of these effects depend on the shape of the window function, which must be suitably chosen to minimize the artefacts while keeping a good accuracy on $k$ and $A$. We found that a convenient choice is the Hamming window~\cite{Heinzel}, which strongly suppresses the amplitude of the first side-lobes at the cost of multiplying the width of the main one by two with respect to a rectangular window. As shown in Fig.~\ref{fig:window}, this suppression efficiently erases the artefacts which are clearly visible with a rectangular window.

\begin{figure} 
\includegraphics[width= \linewidth]{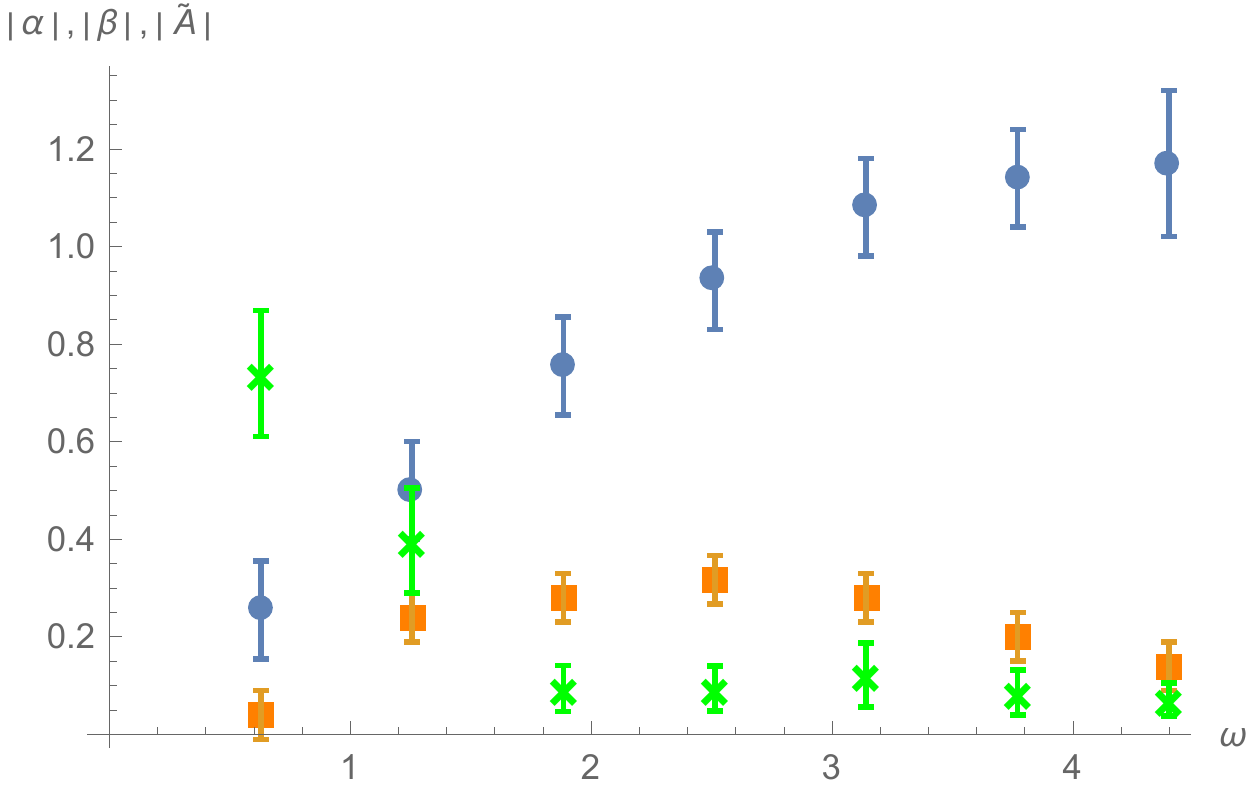} 
\caption{Plots of the norm of the scattering coefficients $| \alpha |$ (blue circles), $| \beta |$ (orange squares), and $|\tilde A|$ (green crosses) observed when sending $I$ waves for seven different frequencies. For $| \alpha |$ and $| \beta |$,  error bars arise from the extension in $k$ space over the finite integration window of $0.95$m. Relative statistical uncertainties are smaller than $10 \%$ and do not contribute significantly. The transmission coefficient $\tilde A$ was obtained by a method similar to that of~\cite{euv15} and averaged over the 80 realizations. Here, error bars show the standard deviation.
}\label{fig:scatcoeffs}
\end{figure}

\subsection{Scattering coefficients} 
When sending a macroscopic coherent wave from the downstream end of the flume, we measured the scattering coefficients for the $B$, $H$ and transmitted modes. Their amplitudes agree with results of numerical simulations only when the static undulation of the background flow is taken into account, as we now describe. 
 
In Fig.~\ref{fig:scatcoeffs} we show the norms of the coefficients $\alpha$, $\beta$, and $\tilde A$ entering in the scattering of the (unit norm) incoming mode $\phi_\omega^I$ into the four outgoing modes~\cite{mic14}, see Fig.~\ref{fig:obstacle}, 
\be \label{eq:scat} 
\phi_\omega^I \to \alpha_\omega  \, \phi_\omega^B + \beta_\omega  \, \phi_\omega^H + A_\omega  \, \phi_\omega^R + \tilde A_\omega  \, \phi_\omega^T.
\ee
The most accurate way we found to measure $\alpha_\omega$ and $\beta_\omega$ consists in using the constructive interferences of the $G_2$ of \eq{eq:G2}:
\begin{eqnarray} 
|\alpha_\omega |
= \frac{G_2(\om, k_I, k_B)}{G_2(\om, k_I, k_I)}
\left\lvert \frac{\partial_\om k_I }{\partial_\om k_B }\right\rvert^{1/2} 
, \nonumber \\ 
| \beta_\omega |
 = \frac{G_2(\om, k_I, k_H)}{G_2(\om, k_I, k_I)} \left\lvert \frac{\partial_\om k_I }{\partial_\om k_H } \right\rvert^{1/2}.
\label{eq:albe}
\end{eqnarray}
In agreement with~\cite{mic14,euv15}, for $\omega \leq \omega_{\rm min}\sim 0.8$Hz, $|\alpha_\omega|$ and $|\beta_\omega|$ 
both significantly decrease whereas the transmission coefficient $|\tilde A_\omega|$ becomes large
(it should reach $1$ when $\omega \to 0$). For $\omega > \omega_{\rm min}$, $|\alpha_\omega|$ increases and saturates to a value close to $1$. $|\beta_\omega|$ also increases and then slowly decreases. In fact, the maximal value of $|\beta_\omega|$ is significantly larger than that we obtained by solving numerically the wave equation, see below. We have not succeeded in measuring the coefficient $A_\omega$ relating long wavelength co-propagating modes. Simulations indicate that it should be smaller than $0.2$. Collecting the data, and assuming that $|A_\omega| \lesssim 0.2 $, the unitarity relation $1= |\alpha_\omega|^2 - |\beta_\omega|^2 + |A_\omega|^2 + |\tilde A_\omega|^2$ expressing the conservation of the norm~\cite{mic14} is obeyed within error bars.

We here summarize the numerical procedure of~\cite{mic14} that we applied in the present case. Under the conditions specified above \eq{disprel}, linear perturbations obey the wave equation \eqref{eq:UCP1}. 
A quartic truncation in $\pd_x$ gives 
\begin{eqnarray} \label{eq:quartic}
\left[ \lp \pd_t + \pd_x U \rp \lp \pd_t + U \pd_x \rp \right.\quad \quad  \quad \quad 
&& \\ \nonumber
 \left. - g \lp \pd_x h \pd_x + \frac{1}{3} \pd_x \lp h \pd_x \rp^3 \rp \right] && \phi = 0,
\end{eqnarray}
where $U$ and $h$ are here $x$-dependent functions describing the background flow and where $\phi$ is the perturbation of the velocity potential at the free surface. It determines the linear variation of the water depth $\delta h$ through $\delta h = -\frac{1}{g}\left(\partial_t+ U \partial_x\right)\phi$.

\begin{figure}
\begin{center}
\includegraphics[width= \linewidth]{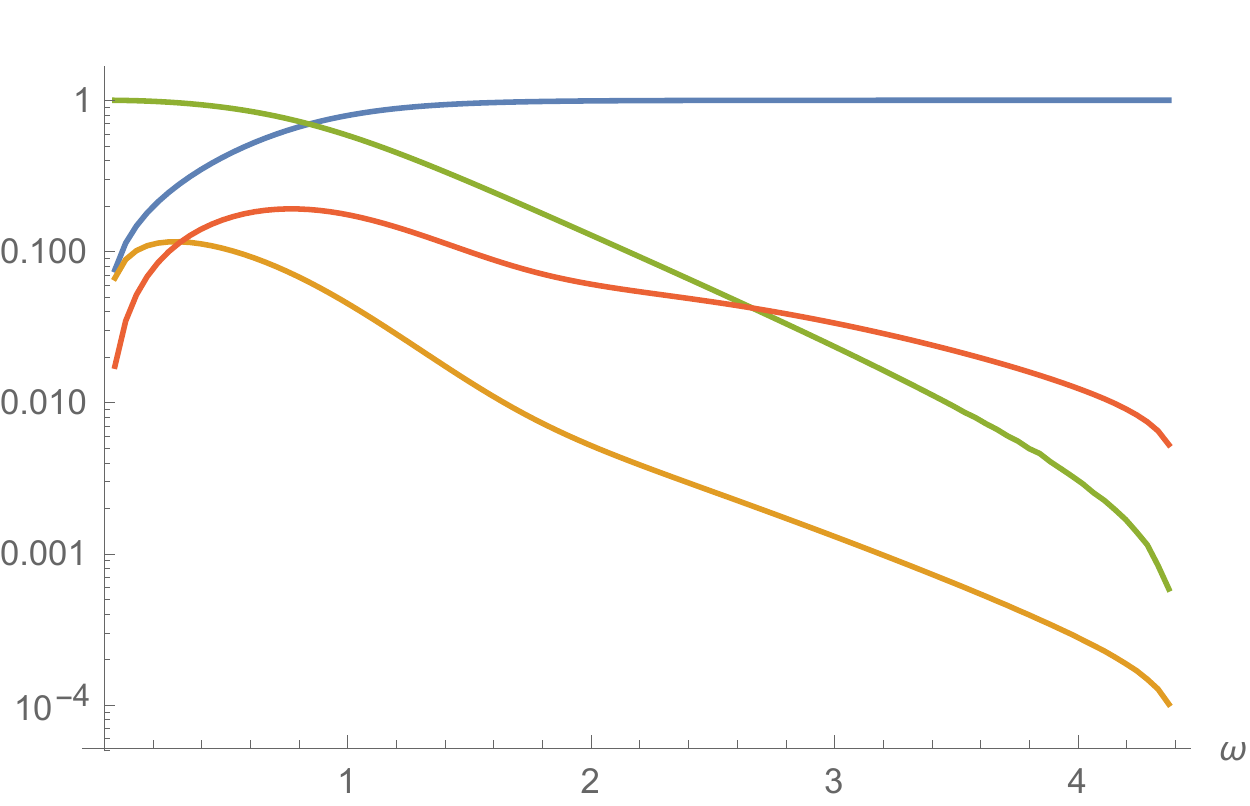}
\end{center}
\caption{In a logarithmic scale, we show the norm of 4 scattering coefficients entering \eq{eq:scat}, namely $\abs{\alpha_\om}$ (blue), $\abs{\beta_\om}$ (orange), $\abs{A_\om}$ (red), and $\abs{\tilde{A_\om}}$ (green) as functions of $\omega$.} \label{fig:hnum}
\end{figure}

We numerically solved \eq{eq:quartic} in a background flow with a water depth of the form 
\begin{eqnarray} \label{eq:hnum}
h(x) = && h_{\rm eff} + (h_0 - h_{\rm eff}) e^{-x^2 / \sigma^2} +  
\frac{A_u}{4} 
\cos \left( k_u x \right)
\\ \nonumber
&&\times \left( 1 + \tanh \left( \frac{x-x_L}{\Delta_L} \right) \right) \left( 1 + \tanh \left( \frac{x-x_R}{\Delta_R} 
 \right) \right), 
\end{eqnarray}
and with $U(x) = q/h(x)$. $q$ and $h_0$ are such that the maximal Froude number matches the observed value $F_{\rm max} \approx 0.85$ and the asymptotic flow velocity is given by its effective value $U_{\rm eff} = 0.37 \rm{m s^{-1}}$. The parameters $x_{R/L}$ and $\Delta_{R/L}$ specify the locations of the two ends of the undulation and their steepness. Its amplitude and wave vector are given by $A_u$ and $k_u$. 

\begin{figure}
\begin{center}
\includegraphics[width= \linewidth]{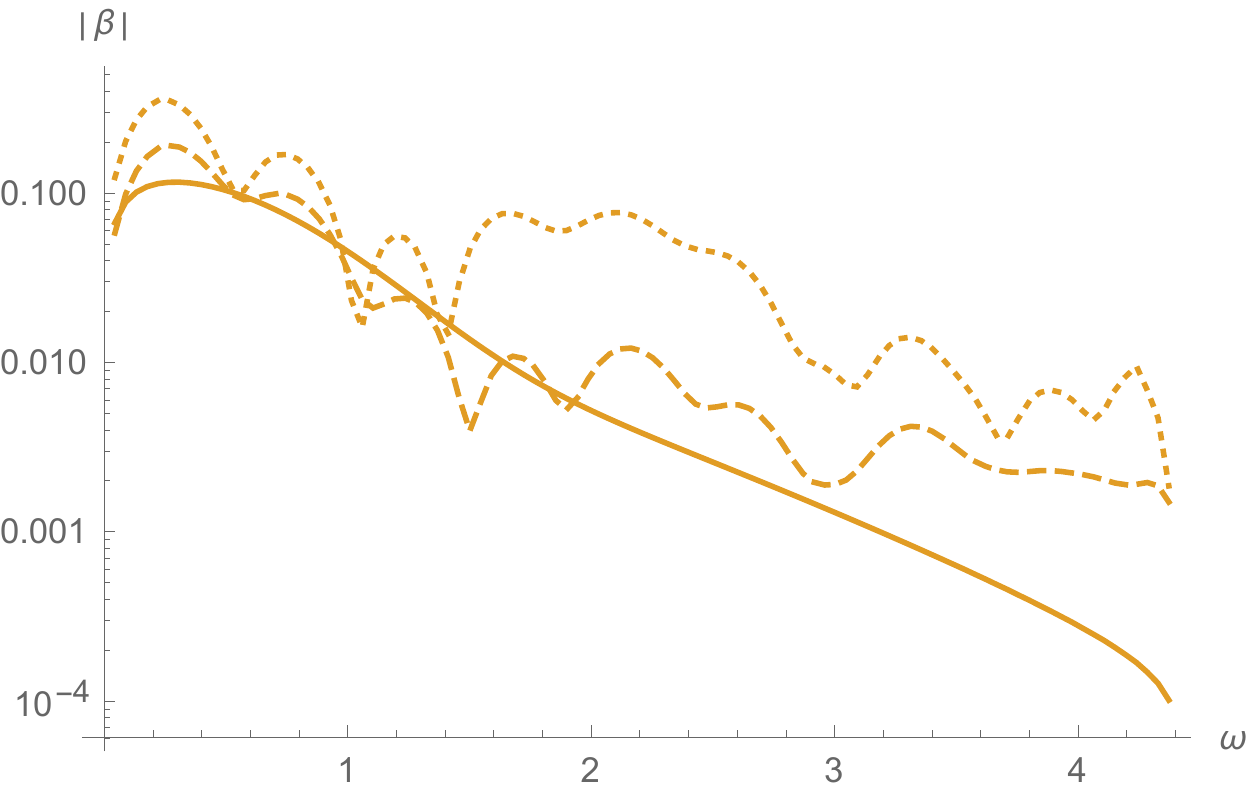}
\end{center}
\caption{We show $\abs{\beta}$ as a function of $\omega$ as a function of $\omega$ for the background flow and for two undulations added on top of it. The corresponding three water depths are given by \eq{eq:hnum} with $A_u = 0$ (continuous), $A_u = 3 {\rm mm}$ (dashed), and $A_u = 9 {\rm mm}$ (dotted). One clearly sees the significant increase of the norm of $\beta_\omega$ when including the scattering on the undulation. 
}\label{fig:bundul}
\end{figure}

We first considered the case without undulation $A_u = 0$, see \fig{fig:hnum} where the logarithms of the norms of the four coefficients entering \eq{eq:scat} are represented as functions of the angular frequency. Because of the quartic truncation of the dispersion relation, the values of $\ommax$ and $\ommin$ slightly differ from the actual ones: $\ommax^{\rm quart} \approx 4.4$~Hz and $\ommin^{\rm quart} \approx 0.72$~Hz. 
We see that the transition from transmission below $\ommin$ to blocking above $\ommin$ reproduces rather well what is found in \fig{fig:scatcoeffs}, in agreement with the analysis of~\cite{euv15}. In fact the main difference concerns the maximal value of the norm of $\beta_\omega$ and its decay for $\omega > \ommin$. 
For instance, for $\om = 3 \ommin^{\rm quart}$, the numerical prediction of $\vert \beta \vert$ is $0.004$, whereas the observed value is $\vert \beta \vert \sim 0.3$.
We believe that this large discrepancy will still be found when replacing the quartic wave equation by a more accurate one. Hence the scattering on a flow without undulation does not seem to reproduce the observed values of $\abs{\beta_\omega}$. 

This conclusion is reinforced by computing $\abs{\beta_\omega}$ on background flows containing an undulation with similar properties as those we observed, see the red curve in \fig{fig:obstacle}. With more precision, the undulation is attached on the downstream side of the obstacle, i.e. $x_L \sim 0$, it has about 
15 oscillations ($k_u x_R \approx 14.3$)
and is slowly damped with $k_u \Delta_R \sim 9$. The wave vector $k_u$ in \eq{eq:hnum} has been chosen to match the zero-frequency root of the quartic wave equation, see \eq{eq:quartic}. Its peak-to-peak amplitude is $2A_u = 6 {\rm mm}$ and $18 {\rm mm}$. When including the scattering on each of these two undulations, we clearly see that the maximum value of $\vert \beta_\om \vert$ is much larger, and its decrease for large $\omega$ slower. We therefore conjecture that the scattering on the undulation plays a significant role in the properties of the coefficients represented in \fig{fig:scatcoeffs}, and in the strength of the correlations presented in \fig{fig:cij}.

\subsection{Nonlinear effects}

\begin{figure}
\begin{center}
\includegraphics[width= \linewidth]{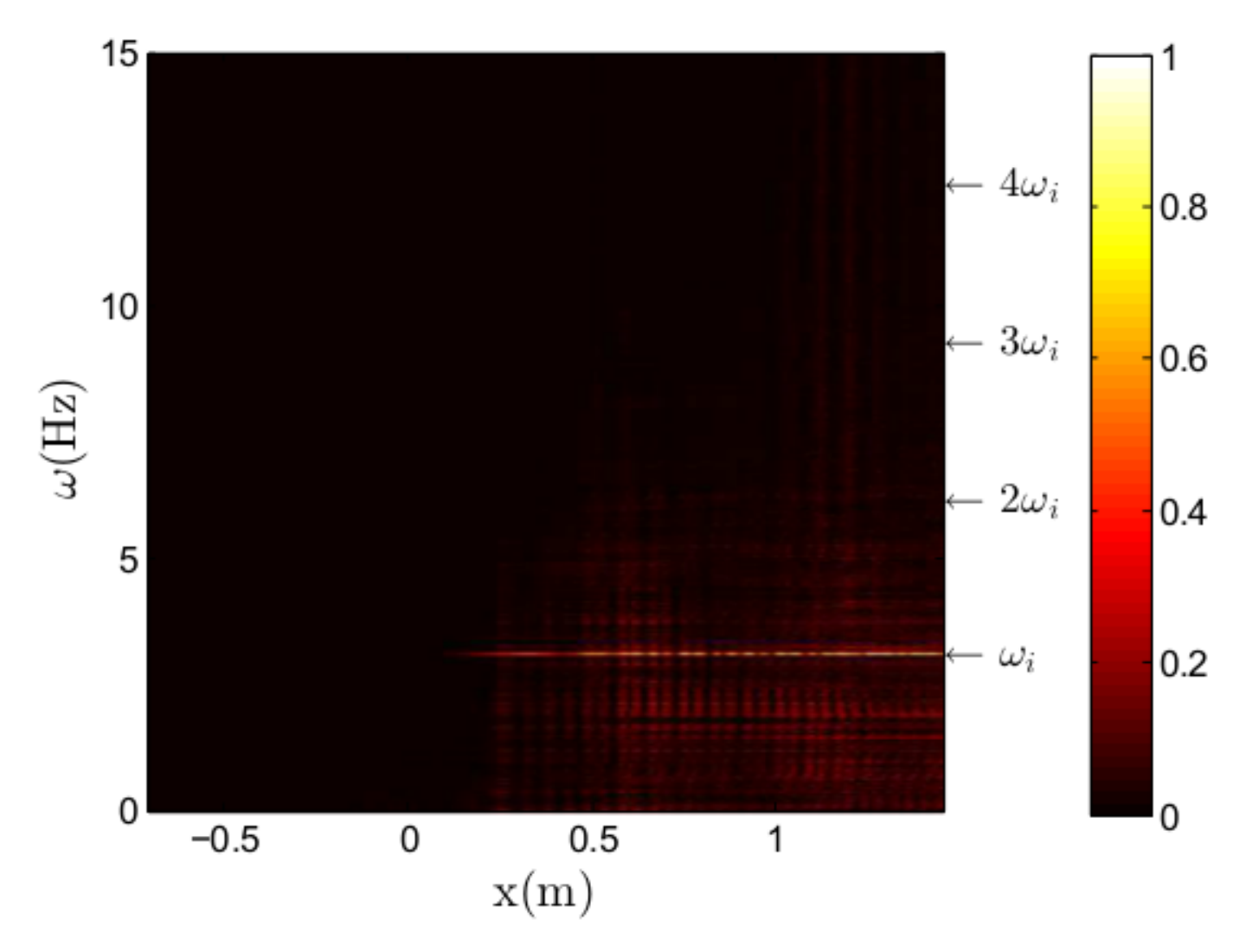}
\includegraphics[width= \linewidth]{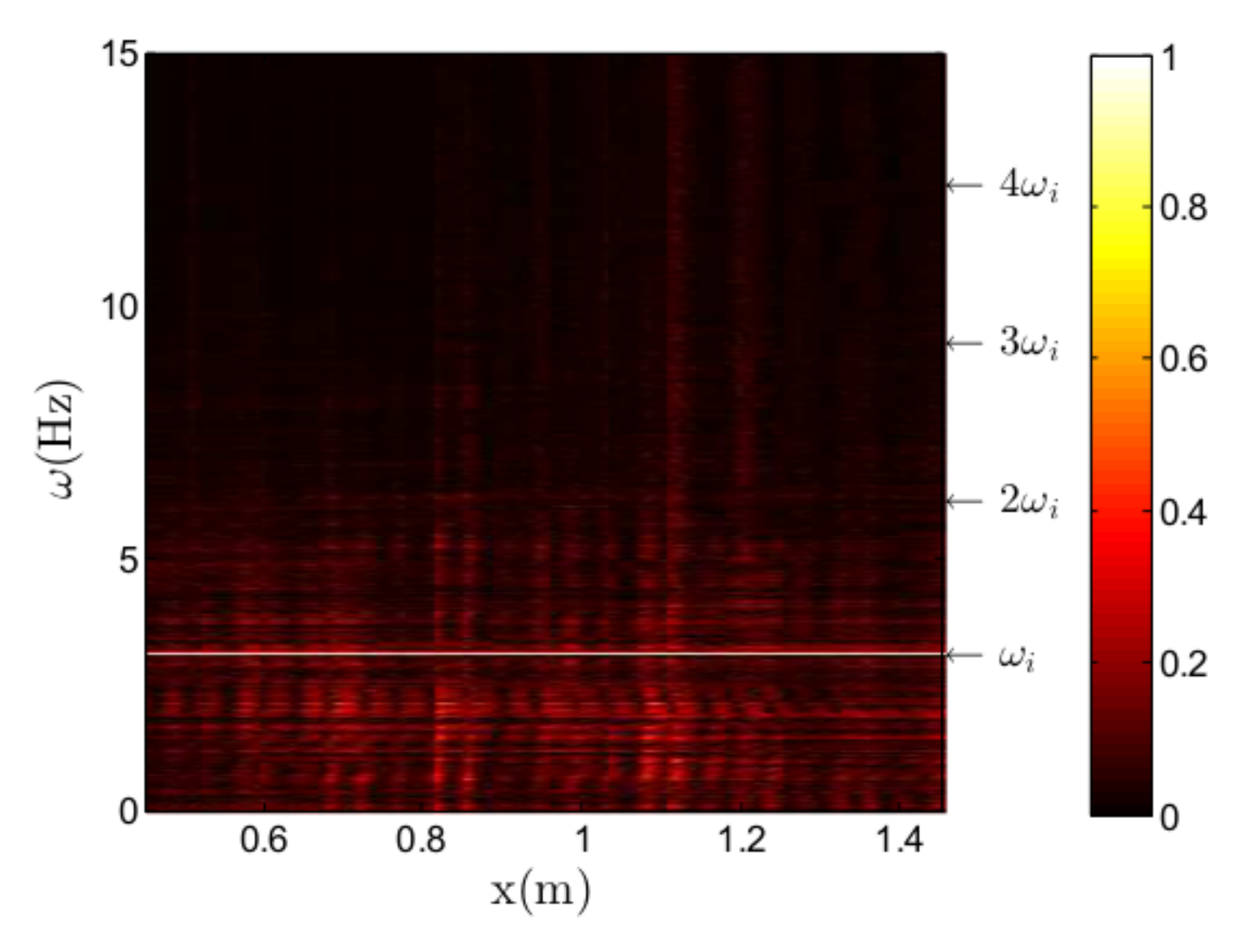}
\end{center}
\caption{Top:  Absolute value of the time Fourier transform $\vert \delta\tilde{h}(\omega,x) \vert$ for the stimulated case with the incoming frequency $\omega_i=3.14$Hz (divided by its maximum value). Bottom: Signal-to-noise ratio $\vert \delta\tilde{h}(\omega,x)\vert \vert /\delta\tilde{h}(\omega_i,x)\vert $ against the space position (in the studied region).
}\label{fig:harmo}
\end{figure}

We now estimate the contribution of nonlinearities in the stimulated case. To first order, their effect is to convert part of the incident waves to harmonics with frequencies which are multiples of that of the wave-maker, and/or to induce a spatial dependence of the frequency~\cite{Herbers}. To obtain the evolution of the frequency content in space, 
we compute the Fourier transform in time $\delta\tilde{h}(\omega,x)$ for each position. 
Results are shown in \fig{fig:harmo} for the incoming frequency $\omega_i=3.14$Hz. 
It clearly shows the absence of instability (increase/decrease of the incoming frequency in space) or harmonics generation.
For instance, the ratio between the signal of the incoming frequency and the first harmonic is the same than the signal-to-noise ratio. We have checked that it remains true for all incident frequencies show in \fig{fig:scatcoeffs}. These results indicate that nonlinear effects are negligible. Since the typical amplitude of the noise fluctuations is ten times smaller than that of the waves sent by the wave-maker, nonlinear effects should a fortiori play no significant role in the noise. 

\subsection{Correlations in a flow without obstacle} 

To verify that the correlations we observed are mainly due to the scattering on the flow inhomogeneities induced by the obstacle, we now briefly study the properties of the noise for a flow over a flat bottom. 
For these measurements, we used three cameras (Point Grey Research Grasshopper3 $2048 \times 2048$) which provide a $2.30$m wide visualization window with a resolution of $0.37$mm/px. 
The spatial window used to compute the Fourier transform is $1 {\rm m}$ long and the experiment lasted $102 {\rm s}$, separated into $10$ intervals of equal duration to compute the averages.

\begin{figure}
\centering
\includegraphics[width=1.\linewidth]{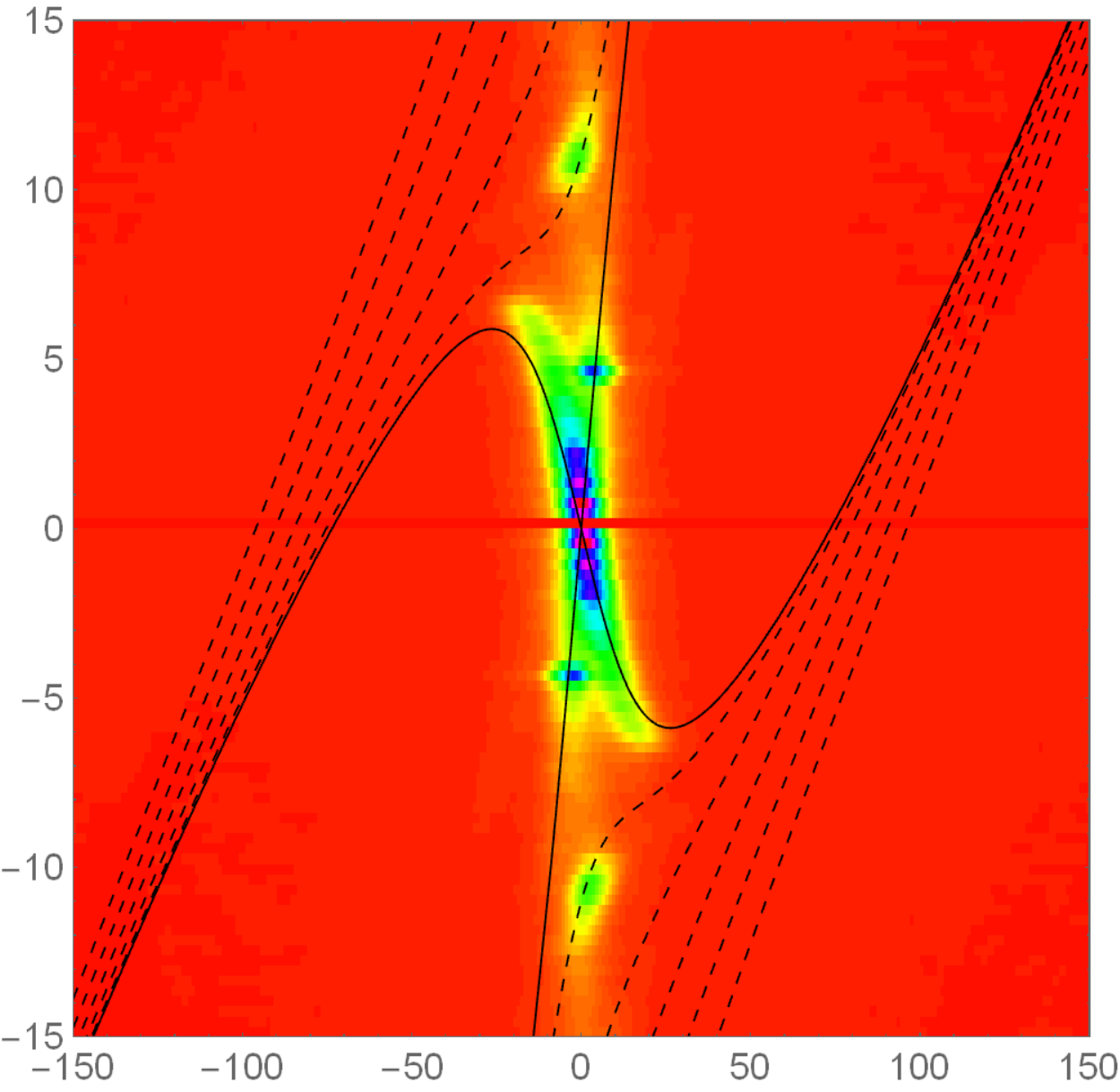} 
\caption{Power spectrum for the flow without obstacle. As in \fig{fig:dispersion}, we show the square root of $\mathcal{P}$, and use the same normalization. 
}
\label{fig:powspec10}
\end{figure}

The power spectrum of the noise in the (nearly homogeneous) flow is shown in \fig{fig:powspec10}. When comparing it with Fig.~1 of the main text, the principal difference concerns the power of the dispersive branches, which is strongly reduced. This is a first indication that the mode conversion is much smaller than in the case with obstacle. The correlation maps at fixed $\omega$ in the $k-k'$ plane show no clear cross-correlations. (This is why we do not show them.)

To quantify the low intensity of the cross-correlations, we estimated the ratios $\abs{c_{\rm IB}} / n_{\rm I}$ and $\abs{c_{\rm IH}} / n_{\rm I}$ for several frequencies, where
\begin{align*}
c_{ab}(\om) \equiv \frac{G_2 \lp \om ; k_a(\om), k_b(\om) \rp}{\sqrt{\frac{d k_a}{d \omega}\frac{d k_b}{d \omega}}},
\end{align*}
and where $n_I$ is the normalized power spectrum of the incoming hydrodynamical I modes.  Although the data seem too noisy to determine a precise value, we found that these two quantities remain smaller than $10^{-3}$. 
By comparison, for the flow with obstacle these quantities are slightly larger than $10^{-2}$ for $\om = 1 {\rm Hz}$ and larger than $0.1$ for larger frequencies. This strongly suggests that most of the observed correlations presented 
in the main text come from the inhomogeneities of the flow due to the obstacle, and the blocking of incoming waves. 

\subsection{Classical and quantum correlations}

It is interesting to compare the techniques we used to those recently used by J. Steinhauer in his study of density fluctuations in a condensed atom gas~\cite{ste15}. Before doing so, it is worth comparing the two systems. 

In both cases, one studies linear perturbations propagating on top of an inhomogeneous flow. The dispersion relation 
is superluminal (anomalous) in~\cite{ste15} as the group velocity in the frame of the fluid increases when increasing the wave number $k$, whereas it is subluminal in our case. The main consequence of this difference appears at the level of the characteristics of the corresponding wave equations. In the superluminal case, the turning point (where the wave is blocked) is located in the supersonic region~\cite{mic16}, whereas in our case it is located in the subsonic region~\cite{mic14}. However, the scattering coefficients are similar between the two cases. Indeed, as explained in~\cite{cou12} there exists a map between them which preserves their main properties in the weak dispersive limit.  
 
More important differences appear at the level of the background flows. 
Firstly, J. Steinhauer created a (nearly) stationary flow which is analogous to a black hole (as the flow velocity increases along the direction of the flow) whereas we dealt with a stationary flow analogous to a white hole (in the downstream region). 
As a result, in our case, the scattered waves are described by the high wave number roots $k_B$ and $k_H$ which both propagate against the flow on the same side of the flow, whereas in~\cite{ste15}
the scattered waves are described by low wave-number roots which propagate on either side of the horizon. Secondly, the flow created by J. Steinhauer is clearly trans-critical as the Mach number $M = v/c$ on the supersonic side is of the order of $4$ whereas our 
flow is sub-critical since $F_{\rm max} \approx 0.85$. Hence, in our case, only waves with frequency above $\om_{\rm min}$ of \fig{fig:maxmin} are essentially blocked, whereas there is no such critical frequency in~\cite{ste15}. 
There is yet another difference which concerns the number of modes involved in the scattering at fixed $\omega$. In our case, 
as can be seen in \eq{eq:scat}, the scattering involves four modes. In his case instead, because the flow is transcritical and monotonic, there are only three of them~\cite{Macher:2009nz}. Finally, our white hole flow is modulated on the downstream region by a zero-frequency undulation which induces some extra scattering, as we discussed above. 

Besides these differences, the important common fact is the stationary linear mode mixing which involves negative energy waves. Hence in both cases there is a steady production of correlated pairs of modes  
propagating away from the horizon in the asymptotic uniform regions. Moreover, in both cases one is dealing with a statistical ensemble of density perturbations. In~\cite{ste15}, the phonon state is described by an ultra low temperature quantum state, 
apparently very close to the vacuum. In our case, we have a random distribution of incoming perturbations. Hence in both cases the appropriate tool to study the effects of the scattering is to consider the two-point function of \eq{eq:G2}, 
parametrized by the wave numbers $k$ and $k'$ measured in the asymptotic homogeneous regions. 
Then, as explained in the text, since the background flow and the probability distribution are both stationary, the cross-correlations should only involve pairs of modes with roots $k_a(\omega), k_b(\omega)$ sharing the same frequency $\omega$. 
At this level, there is an important difference between observations of cold gases and surface waves. 
In the present work, we measured the perturbations as a function of time. From this we could extract the two-point function at two different times, and therefore its behavior at fixed $\omega$. In~\cite{ste15} instead, in situ density perturbations are only measured at one time and, to obtain statistically relevant information, the measurements are repeated many times (always after the same lapse since the formation of the sonic horizon). In that case, only the equal-time correlation function is obtained. As a result, only the integral over $\omega$ of \eq{eq:G2}~\cite{mic16} is obtained, something which could induce a certain blurring of the cross-correlations.

Irrespectively of this difference, the key information concerns the relative importance of the strength of the correlation between the modes $k_a(\omega)$ and $k_b(\omega)$ with respect to their power spectrum (their auto-correlations). This comparison can be made precise by computing the ratio of \eq{eq:g2om}. When dealing with a classical ensemble of perturbations, this ratio is necessarily smaller than 1, as this is guaranteed by a Cauchy-Schwarz inequality. In classical terms, the value of $g_2(\omega; a,b)$ gives the fraction of modes $k_a(\omega)$ and $k_b(\omega)$ that are correlated to each other. 
In quantum settings, the situation is more subtle. When using normal ordered  operators to compute the auto-correlations, (i.e., the mean occupation number of quasi-particles), $g_2(\omega; a,b)$ can be larger than 1, or equivalently, the difference $\Delta = G_2(\omega ; k_a,k_a)\times G_2(\omega ; k_b,k_b) - |G_2(\omega ; k_a,k_b)|^2$ can be negative for a small subset of (entangled) states. In this case, the bi-partite state describing the modes $k_a(\omega)$ and $k_b(\omega)$ is necessarily ``non-separable'', see~\cite{deNova:2015nsa,Busch:2014bza,Boiron:2014npa} for a presentation of these notions and their implementation in the context of the analog Hawking radiation.  

\end{document}